%% file: crowd_hierarchies_kdd.tex
\documentclass{sig-alternate}
\usepackage{graphicx}
\usepackage{color}
\usepackage{balance}  
\usepackage{times}
\usepackage{url}
\usepackage{algorithm}
\usepackage[noend]{algorithmic}
\usepackage{subfigure}
\usepackage{xspace}
\usepackage[noend]{algorithmic}
\usepackage{enumerate}
\usepackage{multirow}
\usepackage{epstopdf}
\usepackage{cleveref}
\usepackage{soul}
\usepackage[font={small,it}]{caption}

\newcommand{\squishlist}{
   \begin{list}{$\bullet$}
    {
      \setlength{\itemsep}{0pt}
      \setlength{\parsep}{3pt}
      \setlength{\topsep}{3pt}
      \setlength{\partopsep}{0pt}
      \setlength{\leftmargin}{1.5em}
      \setlength{\labelwidth}{1em}
      \setlength{\labelsep}{0.5em} } }

\newcommand{\squishend}{
    \end{list}  }

\makeatletter
\def\@copyrightspace{\relax}
\makeatother

\newcommand{\eat}[1]{}

\newcommand{\stitle}[1]{\vspace{0.5em}\noindent\textbf{#1}}

\newtheorem{example}{Example}
\newtheorem{theorem}{Theorem}
\newtheorem{lemma}{Lemma}

\newtheorem{problem}{Problem}

\newcommand{\domain}{\mathcal{D}}
\newcommand{\attributes}{\mathcal{A}_D}
\newcommand{\hierarchy}{\mathcal{H}_D}

\newcommand{\uentities}{\mathcal{E}}

\newif\ifpaper

\newif\iftr
\trtrue 

\begin{document}


\title{CrowdGather: Entity Extraction over Structured Domains}

\numberofauthors{3} 

\author{
	\alignauthor Theodoros Rekatsinas\\
            \affaddr{University of Maryland, College Park} 
                \email{thodrek@cs.umd.edu}
            \alignauthor Amol Deshpande\\
            \affaddr{University of Maryland, College Park} 
                \email{amol@cs.umd.edu}
            \alignauthor Aditya Parameswaran \\
            \affaddr{University of Illinois, Urbana-Champaign} 
                \email{adityagp@illinois.edu}
}

\maketitle

\begin{abstract}
Crowdsourced entity extraction is often used to acquire data for many applications, including recommendation systems, construction of aggregated listings and directories, and knowledge base construction. Current solutions focus on entity extraction using a single query, e.g., only using ``give me another restaurant'', when assembling a list of all restaurants. Due to the cost of human labor, solutions that focus on a single query can be highly impractical.

In this paper, we leverage the fact that entity extraction often focuses on {\em structured domains}, i.e., domains that are described by a collection of attributes, each potentially exhibiting hierarchical structure. Given such a domain, we enable a richer space of queries, e.g., ``give me another Moroccan restaurant in Manhattan that does takeout''. Naturally, enabling a richer space of queries comes with a host of issues, especially since many queries return empty answers. We develop new statistical tools that enable us to reason about the gain of issuing {\em additional queries} given little to no information, and show how we can exploit the overlaps across the results of queries for different points of the data domain to obtain accurate estimates of the gain. We cast the problem of {\em budgeted entity extraction} over large domains as an adaptive optimization problem that seeks to maximize the number of extracted entities, while minimizing the overall extraction costs. We evaluate our techniques with experiments on both synthetic and real-world datasets, demonstrating a yield of up to 4X over competing approaches for the same budget.
\end{abstract}

\input{sections/intro}

\input{sections/prelims}
\input{sections/gainestimators}
\input{sections/solving}
\input{sections/exps}

\input{sections/related}
\input{sections/conclusions}
\balance

\end{document}

%% file: sections/intro.tex


\section{Introduction}
\label{sec:intro}
Combining human computation with traditional computation, commonly referred to as {\em crowdsourcing}, has been recently proven beneficial in extracting knowledge and acquiring data for many application domains, including recommendation systems~\cite{amsterdamer:2014}, knowledge base completion~\cite{kondredi:2014}, entity extraction and structured data collection~\cite{park:2014,trushkowsky:2013}. In fact, extracting information, and entities in particular, from the crowd has been shown to provide access to more fine-grained information that may belong to the long tail of the web or even be completely unavailable on the web~\cite{franklin:2011, Parameswaran:2012, west:2014}.

A fundamental challenge in crowdsourced entity extraction is reasoning about the completeness of the extracted information. Given a task, e.g., ``extract all restaurants in New York'',  that seeks to extract entities from a specific domain by asking human workers, it is not easy to judge if we have extracted all entities (in this case restaurants). This is because we assume an ``open world''~\cite{franklin:2011}.

Recent work~\cite{trushkowsky:2013} has considered the problem of crowdsourced entity extraction using a single type of {\em query} that is asked to humans; for our restaurant case, the query will be ``give me another restaurant in New York''. That paper determines how many times this query must be asked to different human workers before we are sure we have extracted most of the restaurants in New York. However, given the monetary cost inherent in leveraging crowdsourcing, it is easy to see that just using this query repeatedly will not be practical for real-world applications, for two coupled reasons: (a) {\em wasted cost:} we will keep receiving the most popular restaurants and will have to issue many additional queries before receiving new or unseen restaurants, thus, increasing the cost; (b) {\em lack of coverage:}  beyond a point all the restaurants we get will already be present in our set of extracted entities --- thus, we may never end up receiving less popular restaurants at all.

In this paper, our goal is to {\em make crowdsourced entity extraction practical}. To do so, we focus on entity extraction over structured domains, i.e., a domain that can be fully described by a collection of attributes, each potentially being hierarchically structured. For example, in our restaurant case, we could have one attribute about location, one about cuisine, and one about whether the restaurant does takeout. Often the structure of domains in practical applications is already known by design. We can then leverage this structure to use a much richer space of queries asked to human workers, considering all combinations of values for each of these attributes, e.g., ``give me another Moroccan restaurant in Manhattan, New York, that does takeout''.  In this manner, we can leverage these {\em specific, targeted queries} to diversify entity extraction and obtain not-so-popular entities as well.

If we view the structured data domain as a \emph{partially ordered set} (poset), then each query can be mapped to a node in the graph describing its topology. Thus, our goal is to traverse the graph corresponding to the input poset by issuing queries corresponding to various nodes, often multiple times at each node. However, the poset describing the domain can be often large, leading to many additional challenges in deciding which queries to issue at any node: (a)~{\em Sparsity:} Many of the nodes in the poset are likely to be empty, i.e., the queries corresponding to those nodes are likely to not have any answers; avoiding asking queries corresponding to these nodes is essential to keep monetary cost low. (b) {\em Interrelationships:} Many of the nodes in the poset are ``coupled'' with one another; for example, the results from a few queries corresponding to ``give me another Moroccan restaurant in Manhattan, New York'' can inform whether issuing queries corresponding to ``give me another Moroccan restaurant in Manhattan, New York, that does takeout'' is useful or not. We elaborate more on these challenges in Section~\ref{sec:challenges} using examples from a real-world scenario.

Previously proposed techniques~\cite{trushkowsky:2013} do not directly apply to the scenario where we are traversing a poset corresponding to this structured data domain, and new techniques are needed. The main limitation of the aforementioned techniques is that they focus on estimating the completeness of a specific query and are agnostic to cost. As a consequence they do not address the problem of deciding which additional queries are \emph{worth} issuing.  To mitigate these shortcomings, one needs to tune the queries that are asked. However, deciding which queries to ask among a large number of possible queries (exponential in the number of attributes describing the input domain) and when and how many times to ask each query, are both critical challenges that need to be addressed. Furthermore, unlike previous work, we focus on the budgeted case, where we are given a budget and we want to maximize the number of retrieved entities; we believe this is a more practical goal, instead of the goal of retrieving all entities.  
\iftr
Our crowdsourced entity extraction techniques can be useful for a variety of entity extraction applications that are naturally coupled to a structured domain, including:
\squishlist
\item A newspaper that wants to collect a list of today's events to be displayed on the events page every day. 
In this case, the structured data domain could include event type (e.g., music concerts vs.~political rallies) or location, among other attributes.
\item A stock trading firm wants to collect a list of stocks that have been mentioned by popular press on the previous day. In this case, the structured data domain could include stock type, popular press article type, or whether the mention was positive or negative, among other attributes. 
\item A real estate expert wants to curate a list of houses available for viewing today. The structured data domain in this case could include the price range, the number of floors, etc.
\item A university wants to find all the faculty candidates on the job market. The structured data domain in this scenario includes the university of the applicant, specialization, and whether they are Ph.D./Postdoc.
\item The PC chair of a new conference wants to find potential reviewers. The domain describing each of the candidates can be characterized by the university or company of the reviewer, expertise, qualifications, and so on. 
\squishend
\fi

\subsection{A Real-World Scenario}
\label{sec:challenges}
To exemplify the aforementioned challenges we review a large-scale real-world scenario where crowdsourcing is used to extract entities. We consider Eventbrite~(\url{www.eventbrite.com}), an online event aggregator, that relies on crowdsourcing to compile a directory of events with detailed information about the location, type, date and category of each event. \iftr Typically, event aggregators are interested in collecting information about diverse events spanning from conferences and music festivals to political rallies across different location, i.e., countries or cities. In particular, \fi Eventbrite collects information about events across different countries in the world. Each country is split into cities and areas across the country. Moreover, events are organized according to their type and topic. The attributes and their corresponding structure are known in advance and are given by the design of the application. \iftr We collected a dataset from Eventbrite spanning over 63 countries that are divided into 1,709 subareas (e.g., states) and 10,739 cities, containing events of 19 different types, such as rallies, tournaments, conferences, conventions, etc. and a time period of 31 days spanning over the months of October and November. \fi

Two of the three dimensions, i.e., location and time, describing the domain of collected events  are hierarchically structured. The poset characterizing the domain can be fully specified if we consider the cross product across the possible values for location, event type and time. For each of the location, time, type dimensions we also consider a special {\em wildcard} value. Taking the cross-product across the possible values of these dimensions results in poset with a total of 8,508,160 nodes containing 57,805 distinct events overall. We point out that the events associated with a node in the poset overlap with the events corresponding to its descendants. First, we demonstrate how the sparsity challenge applies to Eventbrite.
\begin{example}
We plot the number of events for each node in the poset describing the Eventbrite domain. Out of 8,508,160 nodes only 175,068 nodes are associated with events while the remaining have zero events. Figure~\ref{fig:eventbritepop} shows the number of events per node (y-axis is in log-scale). Most of the populated nodes have less than 100 events. Additionally, the most populated nodes of the domain correspond to nodes at the higher levels of the poset. When extracting events from such a sparse domain one needs to carefully decide on the crowdsourced queries to be issued especially if operating under a monetary budget.
\end{example}
\iftr
\begin{figure}
	\begin{center}
	\includegraphics[clip,scale=0.5]{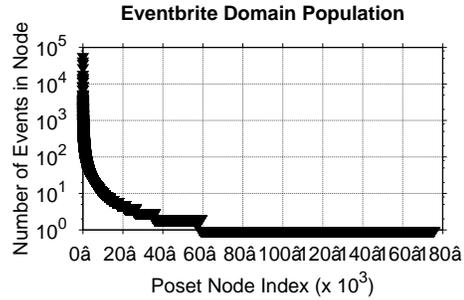}
	\vspace{-10pt}
	\caption{The population of different nodes in the Eventbrite domain.}
	\label{fig:eventbritepop}
	\vspace{-10pt}
	\end{center}
\end{figure}
\fi

As mentioned before, a critical challenge in such large domains is deciding on the queries to ask. However, the hierarchical structure of the data domain presents us with an opportunity. One approach would be to perform a top-down traversal of the poset and issue queries at the different nodes. Nevertheless, this gives rise to a series of challenges: (i) how can one decide on the number of queries to be asked at each node, (ii) when should one progress to deeper levels of the poset and (iii) which subareas should be explored. We elaborate on these in Section~\ref{sec:prelims}. Next, we focus on the second challenge, i.e., the interdependencies across poset nodes. 
\begin{example}
We consider again the Eventbrite dataset and plot the pairwise overlaps of the ten most populous nodes in the domain. Figure~\ref{fig:eventbriteover} shows the Jaccard index for the corresponding node pairs. As shown the event populations corresponding to these nodes overlap significantly. It is easy to see that when issuing queries at a certain domain node, we not only obtain events corresponding to this node but to other nodes in the domain as well.
\end{example}
A critical issue that stems from the overlaps across nodes is being able to decide how many answers to expect when issuing an additional query at a node whose underlying population overlaps with nodes associated with previous queries. In Section~\ref{sec:prelims}, we elaborate more on the dependencies across nodes of the poset.
\iftr
\begin{figure}
	\begin{center}
	\includegraphics[clip,scale=0.5]{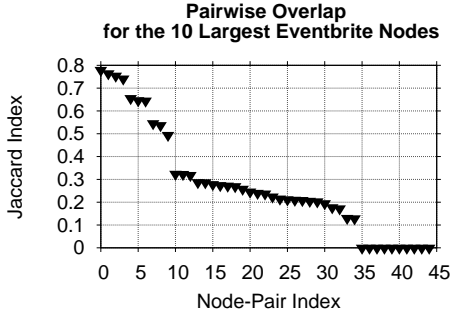}
	\vspace{-10pt}
	\caption{Pairwise overlaps for the 10 most populous nodes.}
	\label{fig:eventbriteover}
	\vspace{-10pt}
	\end{center}
\end{figure}
\fi
\subsection{Contributions}
\label{sec:contributions}
Motivated by the examples above, we study the problem of entity extraction over {\em structured domains}. More precisely, we focus on domains described by a collection of attributes, each following a known {\em hierarchical structure}, i.e., we assume that for each attribute the corresponding hierarchy is known. Such hierarchies are usually dictated by the design of applications. \iftr Moreover, as controlling the overall extraction cost in large-scale applications is crucial we focus on {\em budgeted crowd entity extraction}. \fi

We propose a novel algorithmic framework that exploits the structure of the domain to maximize the number of extracted entities under given budget constraints. In particular, we view the problem of entity extraction as a {\em multi-round adaptive optimization problem}. At  each round we exploit the information on extracted entities obtained by previous queries to adaptively select the crowd query that will maximize the {\em cost-gain} trade-off at each round. The gain of a query is defined as the {\em number of new unique entities extracted}. 

We consider {\em generalized queries} that ask workers to provide us with entities from a domain $D$ and can also include an {\em exclude list}. In general such queries are of the type ``Give me $k$ more entities with attributes $\bar{X}$ that belong in domain $D$ and are not in $\{A, B, ...\}$''. Extending techniques from the species estimation and building upon the multi-armed bandits literature, we introduce a new methodology for estimating the gain for such generalized queries and show how the hierarchical structure of the domain can be exploited to increase the number of extracted entities. Our main contributions are as follows:

\squishlist
\item We study the challenge of information flow across entity extraction queries for overlapping parts of the data domain.
\item We formalize the notion of an exclude list for crowdsourced entity extraction queries and show how previously proposed gain estimators can be extended to handle such queries.
\item We develop a new technique to estimate the gain of generalized entity extraction queries under the presence of little information, i.e., only when a small portion of the underlying entity population has been observed. We empirically demonstrate its effectiveness when extracting entities from sparse domains.
\item We introduce an adaptive optimization algorithm that takes as input the gain estimates for different types of queries and identifies querying policies that maximize the total number of retrieved entities under given budget constraints. 
\item Finally, we show that our techniques can effectively solve the problem of budgeted crowd entity extraction for large data domains on both real-world and synthetic data.
\squishend

%% file: sections/prelims.tex

\section{Preliminaries}
\label{sec:prelims}
In this section we first define structured domains, then describe entities and entity extraction queries or interfaces, along with the response and cost model for these queries. Then, we define the problem of {\em crowd entity extraction} over {\em structured domains} that seeks to maximize the number of extracted entities under budget constraints and present an overview of our proposed framework.

\subsection{Structured Data Domain}
\label{sec:data-domain}

\begin{figure}[h]
\vspace{-10pt}
	\begin{center}
	\includegraphics[clip,scale=0.25]{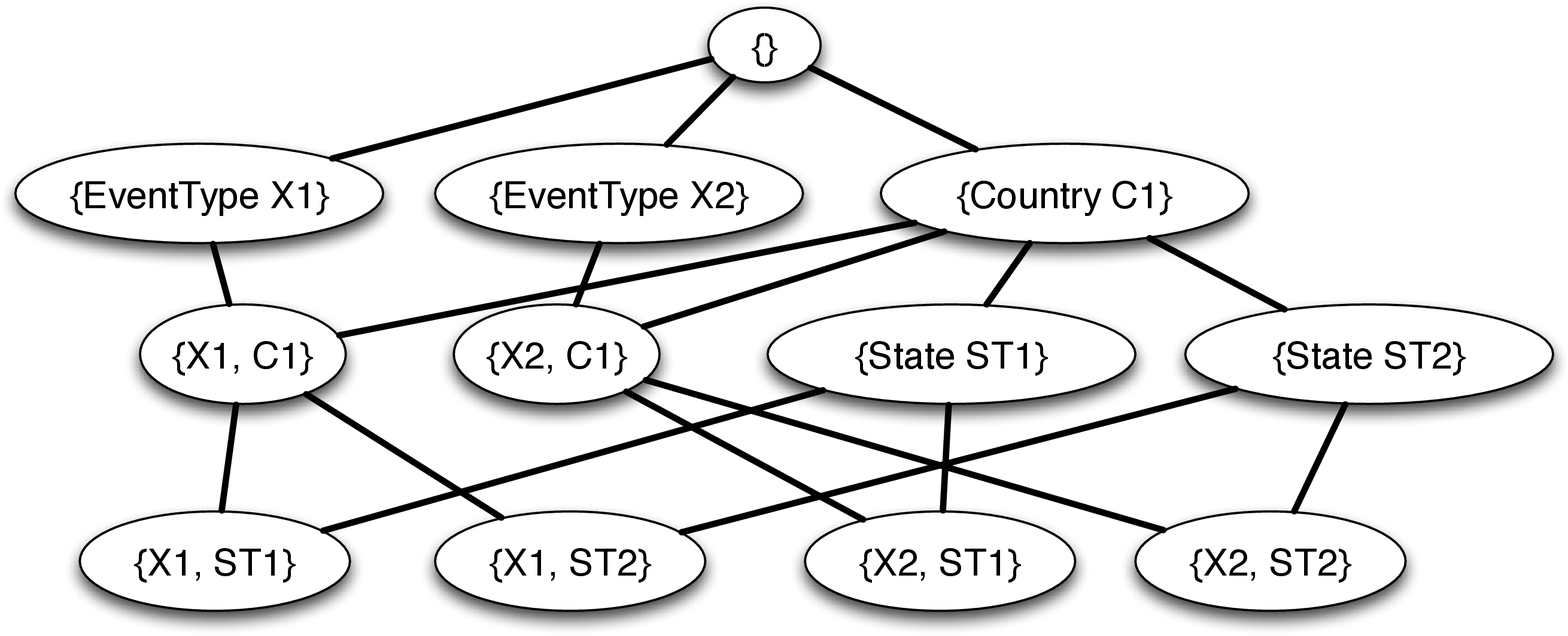}
	\vspace{-10pt}
	\caption{Part of the poset defining the entity domain for Eventbrite.}
	\label{fig:eventslattice}
	\end{center}
	\vspace{-20pt}
\end{figure}

Let $\domain$ be a data domain described by a set of discrete attributes $\attributes = \{A_1, A_2, \dots, A_d\}$. Let $dom(A_i)$ denote the domain of each attribute $A_i  \in \attributes$. We focus on domains where each attribute $A_i$ is hierarchically organized. \iftr For example, consider the Eventbrite domain introduced in Section~\ref{sec:challenges}. The data domain $\domain$ corresponds to all events and the attributes describing the entities in $\domain$ are $\attributes = \{$``Event Type'', ``Location'', ``Date''$\}$. Figure~\ref{fig:eventsdomain} shows the hierarchical organization of each attribute.

\begin{figure}[h]
	\begin{center}
	\includegraphics[clip,scale=0.33]{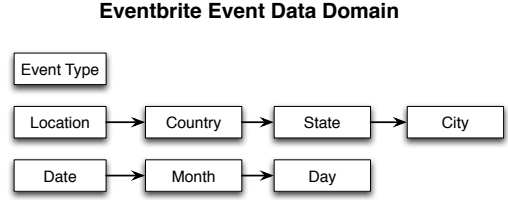}
	\vspace{-10pt}
	\caption{The attributes describing the Eventbrite domain and the hierarchical structure of each attribute.}
	\label{fig:eventsdomain}
	\end{center}
	\vspace{-20pt}
\end{figure}
\fi
The domain $\domain$ can be viewed as a {\em poset}, i.e., a partially ordered set, corresponding to the cross-product of all available hierarchies\footnote{Note that $\domain$ is not a lattice since there is no unique infimum.}. Part of the poset corresponding to the previous example is shown in Figure~\ref{fig:eventslattice}. We denote this cross-product as $\hierarchy$. As can be seen in Figure~\ref{fig:eventslattice}, there are nodes, such as $\{\}$, where no attributes are specified, and nodes, such as $\{X1\}$ and $\{C1\}$ where just one of the attribute values is specified, as well as nodes, such as $\{X2, ST2\}$, where multiple attribute values are specified.

\subsection{Entities and Entity Extraction Queries}
\label{sec:queries}

\stitle{Entities.} Our goal is to extract entities that belong to the domain $\domain$. We assume that each entity $e$ can be uniquely associated with one of the leaf nodes in the hierarchy $\hierarchy$; that is, there is a unique set of ``most-specific'' values of $A_1, \ldots, A_d$ for every entity. For example, in Eventbrite, each entity (here, a local event) takes place in a specific city, and on a specific day. Our techniques also work for the case when entities can be associated only with ``higher level'' nodes, but we focus on the former case for simplicity. 

\stitle{Queries.} Next, we describe queries for extracting entities from the crowd. First, a query $q$ is issued at a node $v \in \hierarchy$; that is, a query specifies zero or more attribute values from $A_1, \ldots, A_d$ that are derived from the corresponding values of $v$, implicitly requiring the worker to find entities that match the specified attribute values.  

Given a query issued at a node, there are three different configurations one can use to extract entities from the crowd: The first configuration corresponds to {\em single entity queries} where workers are required to provide ``one more'' entity that matches the specified attribute values mentioned in the query. Considering the Eventbrite example introduced in the previous section, an example of a single entity query would be asking a worker to provide ``a concert in Manhattan, New York''. The second configuration corresponds to {\em queries of size k} where workers are asked to provide up to $k$ distinct entities. Finally, the last configuration corresponds to {\em exclude list queries}. Here,  workers are additionally provided with a list $E$ of $l$ entities that have already been extracted and are required to provide up to $k$ distinct entities that are not present in the exclude list. It is easy to see that the last configuration generalizes the previous two. Therefore, in the remainder of the paper, we will only consider queries using the third configuration. To describe a query, we will use the notation $q(k,E)$ denoting a query of size $k$ accompanied with an exclude list $E$ of length $l$. We will denote the configuration characterizing the query as $(k,l)$. 

\stitle{Query Response.} Given a query $q(k, E)$ issued at a node $v \in \hierarchy$, a human worker gives us $k$ distinct entities that belong to the domain $\domain$, match the specified attribute values mentioned in the query (derived from $v$), and are not present in $E$. Furthermore, the human worker provides us the information for the attributes that are not specified in $q$ for each of the $k$ entities. For example, if our query is ``a concert in Manhattan, New York'', with $k = 1, E = \emptyset$, the human worker gives us one concert in Manhattan, New York, but also gives us the day on which the concert will take place (here, the missing, unspecified attribute). If the query is ``a concert in the US'', with $k = 1, E = \emptyset$, the human worker gives us one concert in the US, but also gives the day on which the concert will take place, as well as the specific city. If less than $k$ entities are present in the underlying population, workers have the flexibility to report either an empty answer or a smaller number of entities (Section~\ref{sec:excludelist}).

While the reader may wonder if getting additional attributes for entities is necessary, note that this information allows us to reason about which all nodes in $\hierarchy$ the entity belongs to; without this, it is difficult to effectively traverse the poset. Furthermore, we find that in most practical applications, it is useful to get the values of the missing attributes to organize and categorize the extracted entities better. Similar query interfaces that ask users to fully specify the attributes of entities have been proposed in recent literature~\cite{quinn:2014}. 

\iftr
Finally, answers are expected to be duplicated across workers, who may also specify or extract an entity incorrectly. Resolving duplicate entities during extraction is crucial as this information is later used to estimate characterize the completeness of extracted entities, and thus, reason about the gain of additional queries.  Extraction errors can be resolved by leveraging the presence of duplicate information and by applying de-duplication and entity resolution techniques. At a high-level one can use an entity resolution or string similarity (e.g., jaccard coefficient) algorithm to identify duplicate entities. Furthermore, the additional attributes for each entity, can be used to further ascertain similarity of entities. We refer the user to Getoor and Machanavajjhala~\cite{getoor:kdd13} for an overview of entity resolution techniques. Finally, standard truth discovery techniques can be used to identify the correct attribute values for entities. Nevertheless entity resolution and truth discovery are orthogonal problems and not the focus of this paper. In our experiments on real datasets, we found that there were no cases where humans introduced errors to the attribute values of extracted entities. Only minor errors (e.g., misspelled entity names) were detected and fixed manually. \fi

\stitle{Query Cost.} In a typical crowdsourcing marketplace, tasks have different costs based on their difficulty. Thus, crowdsourced queries of different difficulties should also exhibit different costs. We assume we are provided with a cost function $c(\cdot)$ that obeys the following properties:  (a) given a query with fixed size its cost should increase as the size of its exclude list is increasing, and (b) given a query with a fixed exclude list size its cost should increase as the number of requested answer increases. These are fixed upfront by the interface-designer based on the amount of work involved.

\subsection{Crowdsourced Entity Extraction}
\label{sec:extraction}
The basic version of {\em crowdsourced entity extraction}~\cite{trushkowsky:2013} seeks to extract entities that belong to $\domain$, by simply using repeated queries at the root node, with $k = 1, E = \emptyset$. When considering large entity domains, one may need to issue a series of entity extraction queries at multiple nodes in  $\hierarchy$ --- often overlapping with each other --- so that the entire domain is covered. Issuing queries at different nodes ensures that the coverage across the domain will be maximized. 

We let $\pi$ denote a {\em querying policy}, i.e., a chain of queries at different nodes in $\hierarchy$. Notice that multiple queries $q(k,E)$ can be issued at the same node. Let $C(\pi)$ denote the overall cost, in terms of monetary cost of a querying policy $\pi$. We define the gain of a querying policy $\pi$ to be the total number of unique entities, denoted by $\uentities(\pi)$ extracted when following policy $\pi$. Thus, there is a natural tradeoff between the gain (i.e., the number of extracted entities) and the cost of policies. 

Here, we require that the user will {\em only} provide a monetary budget $\tau_c$ imposing a constraint on the total cost of a selected querying policy, and optimize over all possible querying policies across different nodes of $\hierarchy$. Our goal is to identify the policy that maximizes the number of retrieved entities under the given budget constraint. More formally, we define the problem of budgeted crowd entity extraction as follows:

\begin{problem}[Budgeted Crowd Entity Extraction] \ \\
Let $\domain$ be a given entity domain and $\tau_c$ a monetary budget on the total cost of issued queries. The Budgeted Crowd Entity Extraction problem seeks to find a querying policy $\pi^*$ using queries $q(k,E)$ over nodes in $\hierarchy$ that maximizes the number of unique entities extracted $\uentities(\pi^*)$ under the constraint $C(\pi^*) \leq \tau_c$.
\end{problem}
The optimal policy not only specifies the nodes at which queries will be executed but also the size and exclude list of each query.

The cost of a querying policy $\pi$ is defined as the total cost of all queries issued by following $\pi$. We have that $C(\pi) = \sum_{q \in \pi} c(q)$ where the cost of each query $q$ is defined according to a cost model specified by the user. Computing the total cost of a policy $\pi$ is easy. However, the gain $\uentities(\pi)$ of a policy $\pi$ is unknown as we do not know in advance the entities corresponding to each node in $\hierarchy$, and hence, needs to be estimated, as we discuss next. 

The problem of budgeted crowd entity extraction is an instance of a generalization of the {\em stochastic knapsack problem}~\cite{kosuch,steinberg} where each item has a deterministic cost (weight) but a stochastic profit. The stochastic knapsack problem is known to be NP-hard and so is the budgeted crowd entity extraction problem.

\subsection{Underlying Query Response Model}
\label{sec:sampling}
To reason about the occurrence of entities as response to specific queries, we need an underlying query response model. Our query response model is based on the notion of {\em popularity}.

\iftr
\stitle{Popularities.} We assume that each underlying entity has a {\em fixed, unknown popularity value} with respect to crowd workers. Given a query $q(1, \emptyset)$, asking for one entity without using an exclude list, the probability that we will get entity $e$ that satisfies the constraints specified by $q$ is nothing but the popularity value of $e$ divided by the popularity value of all entities $e'$ that also satisfy the constraints in $q$. As an example, if there are only two entities $e_1, e_2$ that satisfy the constraints specified by a given query $q_1$, with popularity values $3$ and $2$,
then the probability that we get $e_1$ on issuing a query $q_1(1, \emptyset)$ is 3/5. If an exclude list $E$ is specified, then the probability that we will get an entity $e \notin E$ is the popularity value of $e$ divided by the popularity values of all entities $e' \notin E$ also satisfying the constraints specified by $q$. {\bf We do not assume that all workers follow the same popularity distribution}. Rather the overall popularity distribution can be seen as an average of the popularity distributions across all workers.

Thus, since workers are asked to provide a limited number of entities as response to a query, each entity extraction query can be viewed as taking a random sample from an unknown population of entities. In the rest of the paper, we will refer to the distribution characterizing the popularities of entities in a population of entities as the {\em popularity distribution} of the population. We note that this is equivalent to the underlying assumption in the species estimation literature~\cite{chao:1992} (Section~\ref{sec:gainestimators}).


Then, estimating the gain of a query $q(k,E)$ at a node $v \in \hierarchy$ is equivalent to estimating the number of new entities extracted by taking additional samples from the population of $v$ given all the retrieved entities by past samples associated with node $v$~\cite{trushkowsky:2013}. 

\stitle{Samples for a Node.} When extracting entities,  the retrieved entities for a node $v$ (i.e., the {\em running sample}) may correspond to two different kinds of samples: (i) those that were extracted by considering the {\bf entire population} corresponding to node $v$ (ii) and those that we obtained by sampling {\bf only a part of the population} corresponding to $v$. Samples for a node $v$ can be obtained either by querying node $v$ or by indirect information flowing to $v$ by queries at other nodes. We refer to the latter case as {\em dependencies across queries}. 
\begin{figure}[h]
	\begin{center}
	\vspace{-10pt}
	\includegraphics[clip,scale=0.3]{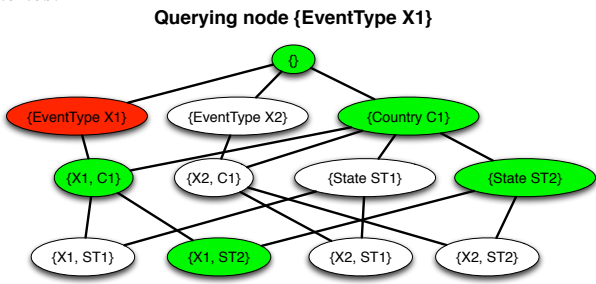}
	\caption{An example query that extract an entity sample from the red node. The nodes marked with green correspond to the nodes for which indirect entity samples are retrieved.}
	\label{fig:query}
	\vspace{-10pt}
	\end{center}
	\vspace{-5pt}
\end{figure}

We use an example considering the poset in Figure~\ref{fig:eventslattice}, to illustrate these two cases. The example is shown in Figure~\ref{fig:query}. Assume a query $q(k,\emptyset)$ issued against node \{EventType X1\}. Assume that the query result contains entities that correspond only to node \{X1,ST2\}. The green nodes in Figure~\ref{fig:query} are nodes for which samples are obtained indirectly without querying them. Notice, that all these nodes are ancestors of \{X1,ST2\}. Analyzing the samples for the different nodes we have:
\squishlist
\item The samples corresponding to nodes \{X1, C1\} and \{X1,ST2\} were obtained by considering their {\em entire population}. The reason is that node \{EventType X1\} is an ancestor of both and the entity population corresponding to it fully contains the populations of both \{X1,C1\} and \{X1,ST1\}. 
\item The samples corresponding to nodes \{ \}, \{Country C1\} and \{State ST2\} were obtained by considering only part of their population. The reason is that the population of node \{EventType X1\} does not fully contain the populations of these nodes. 
\squishend

Samples belonging to both types need to be considered when estimating the gain of a query at a node in $v \in \hierarchy$. To address this issue we merge the extracted entities for each node in $\hierarchy$ into a single sample and treat the unified sample as being extracted from the entire underlying population of the node. As we discuss later in Section~\ref{sec:solving} we develop querying strategies that traverse the poset $\hierarchy$ in a top-down approach, hence, the number of samples belonging in the first category, i.e., samples retrieved considering the entire population of a node, dominates the number of samples retrieved by considering only part of a node's population. Moreover, it has been shown by Hortal et al.~\cite{hortal2006evaluating} that several of the techniques that can be used to estimate the gain of a query (see Section~\ref{sec:gainestimators}) are insensitive to differences in the way the samples are aggregated.
\fi

\subsection{Framework Overview}
\label{sec:framework}
We view the optimization problem described in Section~\ref{sec:extraction} as a multi-round adaptive optimization problem where at each round we solve the following subproblems: 
\squishlist 
\item \textbf{Estimating the Gain for a Query.} For each node in $v \in \hierarchy$, consider the retrieved entities associated with $v$ and estimate the number of new unique entities that will be retrieved if a new query $q(k,E)$ is issued at $v$. This needs to be repeated for different query configurations. 
\item \textbf{Detecting the Optimal Querying Policy.} Using the gain estimates from the previous problem as input, identify the next (query configuration, node) combination so that the total gain across all rounds is maximized with respect to the given budget constraint. When identifying the next query we do not explicitly optimize for the exclude list to be used. We rather optimize for the exclude list size $l$. Once the size is selected, the exclude list is constructed in a randomized fashion. We elaborate more on this design choice in Section~\ref{sec:heuristic}.
\squishend
Our proposed framework iteratively solves the aforementioned problems until the entire budget is used. \iftr Figure~\ref{fig:framework} shows a high-level diagram of our proposed framework.

\begin{figure}
	\begin{center}
	\includegraphics[clip,scale=0.43]{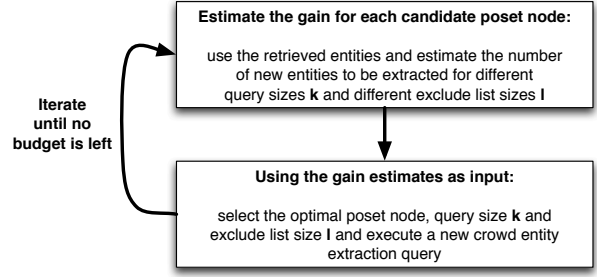}
	\vspace{-10pt}
	\caption{Framework overview for budgeted entity extraction.}
	\label{fig:framework}
	\end{center}
	\vspace{-20pt}
\end{figure}
\fi

%% file: sections/gainestimators.tex

\section{Estimating the Gain of Queries}
\label{sec:gainestimators}
Previous work~\cite{trushkowsky:2013} has drawn connections between this problem and the species estimation literature~\cite{chao:1992}. However, the proposed techniques therein do not work for queries that specify an exclude list. Moreover, they rely on the presence of a relatively large sample and tend to exhibit negative biases~\cite{hwang:2010, shen:2003}, i.e., they underestimate the expected gain. Negative biases can severely impact entity extraction over large domains since nodes that contain entities that belong in the long tail of the popularity distribution may never be queried as they may be deemed to have zero population. In this section, we first review the existing methodology for estimating the gain of a query. Then we discuss how these estimators can be extended to consider an exclude list. Finally, we propose a new gain estimator for generalized queries $q(k,E)$ that exhibits lower biases, and thus, improved performance, in the presence of little information than previous techniques (see Section~\ref{sec:exps}).

\subsection{Previous Estimators}
\label{sec:prevest}
Consider a specific node $v \in \hierarchy$. Prior work only considers samples retrieved from the entire population associated with $v$ and does not consider an exclude list. Let $Q$ be the set of all existing samples retrieved by issuing queries against $v$ without an exclude list. These samples can be combined into a single sample corresponding to multi-set of size $n = \sum_{q \in Q} {\sf size}(q)$. Let $f_i$ denote the number of entities that appear $i$ times in this unified sample, and let $f_0$ denote the number of unseen entities from the population under consideration. Finally, let $C$ be the population coverage of the unified sample. i.e., the fraction of the population covered by the sample  $C = \frac{f_1 + f_2 + ..}{f_0 + f_1 + ...}$.

A new query $q(k,\emptyset)$ at node $v$ can be viewed as increasing the size of the unified sample by $k$. Prior work used techniques from species estimation to estimate the expected number of new entities returned in $q(k,\emptyset)$. Shen et al.~\cite{shen:2003}, derive an estimator for the number of new species $\hat{N}_{Shen}$ that would be found in an increased sample of size $k$. The approach assumes that unobserved entities have equal relative popularity. An estimate of the unique elements found in an increased sample of size $k$ is given by:
\begin{equation}
\label{eq:shen}
\hat{N}_{Shen} = f_0\left( 1 - \left(1 - \frac{1 - C}{f_0}\right)^k\right)
\end{equation}
The second term of Shen's formula corresponds to the probability that at least one unseen entity will be present in a query asking for $k$ more entities. Thus, multiplying this quantity with the number of unseen entities $f_0$ corresponds to the expected number of unseen entities present in the result of a new query $q(k,\emptyset)$.

The quantities $f_0$ and $C$ are unknown and thus need to be estimated considering the entities in the running unified sample. The coverage can be estimated by considering the Good-Turing estimator $\hat{C} = 1 - \frac{f_1}{n}$ for the existing retrieved sample. On the other hand, multiple estimators have been proposed for estimating the number of unseen entities $f_0$. Trushkowsky et al.~\cite{trushkowsky:2013} proposed a variation of an estimator introduced by Chao et al.~\cite{chao:1992} to estimate $f_0$. Nevertheless, the authors argue that the original estimator proposed by Chao performs similarly with their approach when estimating the gain of an additional query $q(k,\emptyset)$. Next, we discuss how one can estimate the return of a query $q(k,E)$ in the presence of an exclude list $E$ of size $l$ and potential negative answers.

\subsection{Exclude Lists and Negative Answers}
\label{sec:excludelist}
A query $q(k, E)$ with $E \ne \emptyset$ issued at node $v \in H_D$ effectively limits the sampling to a restricted subset of the entity population corresponding to node $v$. To estimate the expected return of such a query, we need to update the estimates $\hat{f}_0$ and $\hat{C}$ before applying Equation~(1), by removing the entities in $E$ from the running sample for node $v$ and updating the frequency counts $f_i$ and sample size $n$. This approach requires that the exclude list is known in advance. We discuss how we construct an exclude list in Section~\ref{sec:heuristic}.

Next, we study the effect of {\em negative answers} on estimating the gain of future queries. It is possible to issue a query at a specific node $v \in \hierarchy$ and receive no entities, i.e., we receive a negative answer. This is an indication that the underlying entity population of $v$ is empty. In such a scenario, we assign the expected gain of future queries at $v$ and all its descendants to zero. Another type of negative answer corresponds to issuing a query at an ancestor node $u$ of $v$ and receiving no entities for $v$. In this case, we do not update our estimates for node $u$ as entities from other descendants of $u$ may be more popular than entities associated with $u$.

\subsection{Direct Gain Estimation}
\label{sec:newestim}
The techniques reviewed in Section~\ref{sec:prevest} result in negative bias when the number of observed entities from a population represents only a {\em small fraction} of the entire population~\cite{hwang:2010, shen:2003}. This holds for the large and sparse domains we consider in this paper. To address this problem, Hwang and Shen~\cite{hwang:2010} proposed a regression based technique to estimate $f_0$ and show that it results in smaller biases. However, estimating the total gain of a query requires coupling this new estimator with Equation~(1), thus, it may still exhibit negative bias. To eliminate negative bias, we propose a direct estimator for the gain of generalized queries $q(k,E)$ without using Equation~(1). We build upon the techniques in~\cite{hwang:2010} and use a regression based technique that captures the structural properties of the expected gain function. 

Let $S$ denote the total number of entities in the population under consideration and $p_i$ the abundance probability (i.e., popularity) of entity $i$. Given a sample of size $n$ from the population, define $K(n)$ to be $K(n) = \frac{\sum_{i=1}^S (1-p_i)^n}{\sum_{i=1}^S p_i(1-p_i)^{n-1}}$. First, we focus on queries without an exclude list. Later we relax this and discuss queries with exclude lists. We have the following theorem on query gain:

\vspace{-5pt}\begin{theorem}
\label{newgain}
Given a node $v \in \hierarchy$ and a corresponding entity sample of size $n$, let $f_1$ and $f_2$ denote the number of entities that appear exactly once (i.e., singletons) and exactly twice respectively. Let $G$ denote the number of new items retrieved by a query $q(m,\emptyset)$. We have that:

\vspace{-10pt}\begin{equation}
\label{eq:dirgain}
G = \frac{1}{(1 + \frac{K^{\prime}}{n+m})}(K\frac{f_1}{n} - K^{\prime}\frac{f_1(1-\frac{1}{n}2\frac{f_2}{f_1})^m}{n+m})
\end{equation}
where $K = K(n)$ and $K^{\prime} = K(n+m)$.
\end{theorem}
\iftr
\begin{proof}
To derive the new estimator we make used of the generalized jackknife procedure for species richness estimation~\cite{heltshe1983estimating}. Given two (biased) estimators of $S$, say $\hat{S}_1$ and $\hat{S}_2$, let $R$ be the ratio of their biases:
\begin{equation}
R = \frac{E(\hat{S}_1) - S}{E(\hat{S}_2) - S}
\end{equation}
By the generalized jackknife procedure, we can completely eliminate the bias resulting from either $\hat{S}_1$ or $\hat{S}_2$ via
\begin{equation}
S = G(\hat{S}_1, \hat{S}_2) = \frac{\hat{S}_1 - R\hat{S}_2}{1 - R}
\label{eq:jknife}
\end{equation}
provided the ratio of biases $R$ is known. However, $R$ is unknown and needs to be estimated. 

Let $D_n$ denote the number of unique entities in a unified sample of size $n$. We consider the following two biased estimators of $S$: $\hat{S_1} = D_n$ and $\hat{S}_2 = \sum_{j=1}^n D_{n-1}(j)/n = D_n - f_1/n$ where $D_{n-1}(j)$ is the number of species discovered with the $j$th observation removed from the original sample. Replacing these estimators in Equation~(4) gives us:
\begin{equation}
S = D_n +\frac{R}{1-R}\frac{f_1}{n}
\end{equation}
Similarly, for a sample of increased size $n+m$ we have:
\begin{equation}
S = D_{n+m} +\frac{R^{\prime}}{1-R^{\prime}}\frac{f^{\prime}_1}{n+m}
\end{equation}
where $R^{\prime}$ is the ratio of the biases and $f^{\prime}_1$ the number of singleton entities for the increased sample. Let $K = \frac{R}{1-R}$ and $K^{\prime} = \frac{R^{\prime}}{1-R^{\prime}}$. Taking the difference of the previous two equations we have:
\begin{equation}
D_{n+m} - D_{n} = K\frac{f_1}{n} - K^{\prime}\frac{f^{\prime}_1}{n+m}
\end{equation}
Therefore, we have:
\begin{equation}
\label{eq:new}
G = K\frac{f_1}{n} - K^{\prime}\frac{f^{\prime}_1}{n+m}
\end{equation}
We need to estimate $K$, $K^{\prime}$ and $f^{\prime}_1$. We start with $f^{\prime}_1$, which denotes the number of singleton entities in the increased sample of size $n+m$. Notice, that $f^{\prime}_1$ is not known since we have not obtained the increased sample yet, so we need to express it in terms of $f_1$, i.e., the number of singletons, in the running sample of size $n$. We have:
\begin{equation}
f^{\prime}_1 = G + f_1 - f_1^c
\end{equation}
where $f_1^c$ denotes the number of old singleton entities from the sample of size $n$ that appeared in the additional query of size $m$. Let $E_1$ denote the set of singleton entities in the old sample of size $n$. We approximate $f_1^c$ by its expected value:
\begin{equation}
\hat{f}_1^c = \sum_{e \in E_1} \Pr[\mbox{e appears in query of size $m$}]
\end{equation}
We compute the probability of an old singleton entity appearing in an additional query as follows. Let $p_e$ denote the popularity of entity $e$. As described before, an additional query of size $m$ corresponds to taking a sample of size $m$ from the underlying entity population without replacement. However, $m$ is significantly smaller compared to the size of the underlying population, thus, we can consider a that taking a sample of size $m$ corresponds to taking a sample {\em with replacement}. Following this we have that:
\begin{equation}
\Pr[\mbox{e appears in query of size $m$}] = 1 - (1-p_e)^m
\end{equation}
Following a standard approach in the species estimation literature we assume that the popularity of retrieving a singleton entity again is the same for all singleton entities. This popularity can be computed using the corresponding Good-Turing estimator considering the running sample. We have:
\begin{equation}
\forall e \in E_1, p_e = p_1 = \hat{\theta}(1) = \frac{1}{n}2\frac{f_2}{f_1}
\end{equation}
where $f_2$ is the number of entities that appear twice in the sample and $f_1$ is the number of singletons. 
Eventually we have that:
\begin{equation}
\hat{f}_1^c = f_1(1 - (1-p_1)^m)
\end{equation}
and
\begin{equation}
f^{\prime}_1 = G + f_1(1-p_1)^m
\end{equation}
Replacing the last equation in Equation~(8) we have:
\begin{align}
&G = K\frac{f_1}{n} - K^{\prime}\frac{G + f_1(1-p_1)^m}{n+m} \nonumber \\
&G = K\frac{f_1}{n} - K^{\prime}\frac{G}{n+m} - K^{\prime}\frac{f_1(1- P)}{n+m} \nonumber \\
&G(1 + \frac{K^{\prime}}{n+m}) = K\frac{f_1}{n} - K^{\prime}\frac{f_1(1- P)}{n+m} \nonumber \\
&G = \frac{1}{(1 + \frac{K^{\prime}}{n+m})}(K\frac{f_1}{n} - K^{\prime}\frac{f_1(1-p_1)^m}{n+m}) \nonumber
\end{align}
\end{proof}
\fi
All quantities apart from $K$ and $K^{\prime}$ in Equation~(2) are known. The value of $K$ can be estimated using the regression approach introduced by Hwang and Shen~\cite{hwang:2010}. \iftr From the Cauchy-Schwarz inequality we have that:
\begin{equation}
K = \frac{\sum_{i=1}^S (1-p_i)^n}{\sum_{i=1}^S p_i(1-p_i)^{n-1}} \geq \frac{(n-1)f_1}{2f_2}
\end{equation}
This can be generalized to:
\begin{equation}
K=\frac{nf_0}{f_1} \geq \frac{(n-1)f_1}{2f_2} \geq \frac{(n-2)f_2}{3f_3} \geq \dots
\end{equation}
Let $g(i) = \frac{(n-i)f_i}{(i+1)f_{i+1}}$. From the above we have that the function $g(x)$ is a smooth monotone function for all $x \geq 0$. Moreover, let $y_i$ denote a realization of $g(i)$ mixed with a random error. Hwang and Shen how one can use an exponential regression model to estimate $K$. The proposed model corresponds to:
\begin{equation}
y_i = \beta_0\exp(\beta_1i^{\beta_2}) + \epsilon_i
\end{equation}
where $i = 1, \dots, n-1$, $\beta_0 > 0$, $\beta_1 < 0$, $\beta_2 >0$ and $\epsilon_i$ denotes random errors. It follows that $K = \beta_0$. \fi
To estimate the value of $K^{\prime}$ for an increased sample of size $n+m$, we first show that $K$ increases monotonically as the size of the running sample increases. 

\begin{lemma}
\label{monotonicity}
The function $K(n) = \frac{\sum_{i=1}^S (1-p_i)^n}{\sum_{i=1}^S p_i(1-p_i)^{n-1}}$ increases monotonically, i.e., $K(n+m) \geq K(n), \forall n,m > 0$.
\end{lemma}
\iftr
\begin{proof}
In the remainder of the proof we will denote $K(n+m)$ as $K^{\prime}$. By definition we have that $K = \frac{\sum_{i=1}^S (1-p_i)^n}{\sum_{i=1}^S p_i(1-p_i)^{n-1}}$ and $K^{\prime} = \frac{\sum_{i=1}^S (1-p_i)^{n+m}}{\sum_{i=1}^S p_i(1-p_i)^{n+m-1}}$. We want to show that:

{\small
\begin{align}
&\frac{\sum_{i=1}^S (1-p_i)^{n+m}}{\sum_{i=1}^S p_i(1-p_i)^{n+m-1}} \geq \frac{\sum_{i=1}^S (1-p_i)^n}{\sum_{i=1}^S p_i(1-p_i)^{n-1}} \nonumber \\
&\sum_{i=1}^S (1-p_i)^{n+m}\sum_{j=1}^S p_j(1-p_j)^{n-1} \geq \sum_{i=1}^S p_i(1-p_i)^{n+m-1}\sum_{j=1}^S (1-p_j)^n\nonumber \\
&\sum_{i,j:i\prec j}[(1-p_i)^{n+m}p_j(1-p_j)^{n-1} - p_i(1-p_i)^{n+m-1}(1-p_j)^n + \nonumber \\
& + (1-p_j)^{n+m}p_i(1-p_i)^{n-1} - p_j(1-p_j)^{n+m-1}(1-p_i)^n] \geq 0 \nonumber \\
&\sum_{i,j:i\prec j}[(1-p_i)^{n-1}(1-p_j)^{n-1}(p_j-p_i)((1-p_i)^{m} - (1-p_j)^{m}) \geq 0
\end{align}}

But the last inequality always holds since each term of the summation is positive. In particular, if $p_j \geq p_i$ then
also $1-p_i \geq 1-p_j$ and if $p_j \leq p_i$ then $1-p_i \leq 1-p_j$.
\end{proof}
\fi
Given the monotonicity of function $K$, we model $K$ as a generalized logistic function of the form $K(x) = \frac{A}{1+exp(-G(x-D))}$. As we observe samples of different sizes for different queries we estimate $K$ as described above and therefore we observe different realizations of $f(\cdot)$. Thus, we can learn the parameters of $f$ and use it to estimate $K^{\prime}$. In the presence of an exclude list of size $l$ we follow the approach described in Section~\ref{sec:excludelist} to update the quantities $f_i$ and $n$ used in the analysis above. 

%

%% file: sections/solving.tex

\section{Discovering Querying Policies}
\label{sec:solving}
Next, we focus on the second component of our proposed algorithmic framework and introduce a multi-round adaptive optimization algorithm for identifying querying strategies that maximize the total gain across all rounds under the given budget constraints. We build upon ideas from the multi-armed bandit literature~\cite{Auer:2003,EvenDar06actionelimination}. At each round, the proposed algorithm uses as input the estimated gain or return for different generalized queries $q(k,E)$ at the different nodes in $\hierarchy$. Before presenting our framework we list several challenges associated with this adaptive optimization problem.

\squishlist
\item The first challenge is that the number of nodes in $\hierarchy$ is exponential in the number of attributes $\attributes$ describing the domain of interest. Querying every possible node to estimate its expected return for different queries $q(k,E)$ is prohibitively expensive. That said, typical budgets do not allow algorithms to query all nodes in the hierarchy, so this intractability may not hurt us all that much. For example, we keep estimates for each of the nodes for which at least one entity has been retrieved.
\item The third challenge is balancing the tradeoff between {\em exploitation} and {\em exploration}~\cite{Auer:2003}. The first refers to querying nodes for which sufficient entities have been retrieved and hence we have an accurate estimate for their expected return; the latter refers to exploring new nodes in $\hierarchy$ to avoid locally optimal policies.
\squishend

\subsection{Balancing Exploration and Exploitation}
\label{sec:balancing}
While issuing queries $q(k,E)$ at different nodes of $\hierarchy$ we obtain a collection of entities that can be assigned to different nodes in $\hierarchy$. For each node we can estimate the return of a query $q(k,E)$ using the estimators presented in Section~\ref{sec:gainestimators}. However, this estimate is based on a rather small sample of the underlying population. Thus, exploiting this information at every round may lead to suboptimal decisions. This is the reason why one needs to balance the trade-off between exploiting nodes for which the estimated return is high and nodes that have not been queried many times. Formally, the latter corresponds to upper-bounding the expected return of each potential action with a confidence interval that depends on both the variance of the expected return and the number of times an action has been evaluated.

Let $r(\alpha)$ denote the expected return of action $\alpha$ that is an estimate of the true return $r^*(\alpha)$. Moreover, let $\sigma(\alpha)$ be an error component on the return of action $\alpha$ chosen such that $r(\alpha) - \sigma(\alpha) \leq r^*(\alpha) \leq r(\alpha) + \sigma(\alpha)$ with high probability. The parameter $\sigma(\alpha)$ should take into account both the empirical variance of the expected return as well as our uncertainty if an action or similar actions (e.g., queries with different $k, E$ but at the same node) has been chosen few times. Let $n_{\alpha,t}$ be the number of times we have chosen action $\alpha$ by round $t$, and let $v_{\alpha,t}$ denote the maximum value between some constant $c$ (e.g., $c = 0.01$) and the empirical variance for action $\alpha$ at round $t$. The latter can be computed using bootstrapping over the retrieved sample and applying the estimators presented in Section~\ref{sec:newestim} over these bootstrapped samples. Several techniques have been proposed in the multi-armed bandits literature to compute the parameter $\sigma(\alpha)$~\cite{teytaud:inria-00173263}. Teytaud et al.~\cite{teytaud:inria-00173263} showed that techniques considering both the variance and the number of times an action has been chosen tend to outperform other proposed methods. \iftr Based on this observation, we choose to use the following formula for sigma:
\begin{equation}
\label{eq:upper}
\sigma(\alpha) = \sqrt{\frac{v_{\alpha,t}\cdot\log(t)}{n_{\alpha,t}}}
\end{equation}
\fi

\subsection{A Multi-Round Querying Policy Algorithm}
\label{sec:heuristic}
We now introduce a multi-round algorithm for solving the budgeted entity enumeration problem. At a high-level, the algorithm proceeds as follows: Instead of considering all potential queries $q(k,E)$ that can be issued at the different nodes of $\hierarchy$, we consider all potential query configurations $(k,l)$. In particular, we do not optimize directly for the exclude list to be used in a further query but rather for the size $l$ of it. Once we decide on $l$ the exclude list $E$ can be constructed following a randomized approach, where $l$ of the retrieved entities are included in the list uniformly at random. The generated list can be used to update the frequency counts $f_i$ and sample size $n$ and estimate the gain of the query. Bootstrapping can also be used to obtain improved estimates. 

We follow a randomized approach as a deterministic construction of $E$ that picks the {\em l-most} popular items in the running sample is very sensitive to the {\em observed popularity distribution}. When the number of observed entities corresponds to a small portion of the entire population - as in the scenarios we consider in this paper - the individual entity popularity estimates tend to be very noisy.  We empirically observed that a deterministic construction of a limited size exclude list, especially during early queries, leads to poor popularity estimates. Thus, we choose to follow a randomized approach.

Let $\mathcal{S}$ denote the set of all potential query configurations $(k,l)$ that can be issued at the different nodes of $\hierarchy$ during a round $r$. Moreover, let $r(\alpha) + \sigma(\alpha)$ and $c(\alpha)$ be the upper-bounded return (i.e., gain) and cost for an action $\alpha \in \mathcal{S}$. At each round the algorithm identifies an action in $\mathcal{S}$ that maximizes the quantity $\frac{r(\alpha) + \sigma(\alpha)}{c(\alpha)}$ under the constraint that the cost of action $\alpha$ is less or equal to the remaining budget. Since we are operating under a specified budget one can view the problem in hand as a variation of the typical knapsack problem. If no such action exists then the algorithm terminates. Otherwise the algorithm issues the query corresponding to action $\alpha$, updates the set of unique entities obtained from the queries, the remaining budget and updates the set of potential queries that can be executed in the next round.  An overview of this algorithm is shown in Algorithm~\ref{algo:overall}. 

As discussed before, the size of $\hierarchy$ is exponential to the values of attributes describing it, and thus, considering all the possible queries for the different nodes of $\hierarchy$ can be prohibitively expensive. Next, we discuss how one can initialize and update the set of potential actions as the algorithm progresses based the structure of the poset $\hierarchy$ and the retrieved entities from previous rounds. 

\begin{algorithm}[h]
\small\caption{Overall Algorithm}
\label{algo:overall}
\begin{algorithmic}[1]
\STATE {\bf Input:} $\hierarchy$: the hierarchy describing the entity domain; $r,\sigma$: value oracle access to gain upper bound; $c$: value oracle access to the query costs; $\beta_c$: query budget;
\STATE {\bf Output:} $\uentities$: a set of extracted distinct entities;
\STATE $\uentities \leftarrow \{\}$
\STATE $RB \leftarrow \beta_c$ /* Initialize remaining budget */
\STATE $\mathcal{S} \leftarrow$ {\sf UpdateActionSet($\hierarchy$)}
\WHILE {$RB > 0$ and $S \neq \{\}$}
	\STATE $\alpha \leftarrow \arg\max_{\alpha \in {\mathcal{S}}} \frac{r(\alpha)+\sigma(\alpha)}{c(\alpha)}$ such that $RB - c(\alpha) >0$
	\IF {$\alpha$ is NULL }
		\STATE break;
	\ENDIF
	\STATE $RB \leftarrow RB - c(\alpha)$ /* Update budget */
	\STATE Issue query corresponding to $\alpha$
	\STATE $E \leftarrow$ entities from query
	\STATE $\uentities \leftarrow \uentities \cup E$ /* Update unique entities */
	\STATE $\mathcal{S} \leftarrow$ {\sf UpdateActionSet($\hierarchy$)}
\ENDWHILE
\RETURN $\uentities$
\end{algorithmic}
\end{algorithm}

\subsection{Updating the Set of Actions}
Due to the exponential size of the poset $\hierarchy$, we need to limit the set of possible actions Algorithm~\ref{algo:overall} considers by exploiting the structure the given domain $\hierarchy$. We propose an algorithm that updates the set of actions by traversing the input poset in a top-down manner and adds new actions that correspond to queries for nodes that are {\em direct descendants} of already queried nodes. Due to the hierarchical structure of the poset nodes at higher levels of the poset correspond to larger populations of entities. Therefore, issuing queries at these nodes can potentially result in a larger number of extracted entities. Also, traversing the poset in a top-down manner allows one to detect sparsely populated areas of the poset.

Our approach for updating the set of available actions (Alg.~\ref{algo:updateactions}) proceeds as follows: If the set of available actions is empty start by considering all possible queries that can be issued at the root of $\hierarchy$ (Ln. 4-5). The set of possible queries corresponds to queries $q(k,E)$ for all combinations of the values of parameters $k$ and $l$. Recall that $E$ is constructed in a randomized fashion once $l$ is determined. Recall that these are pre-specified by the designer of the querying interface. If the set of available actions is not empty, we consider the node associated with the action selected in the last round and populate the set of available actions with all the queries corresponding to its direct descendants (Ln. 7-9), i.e., by traversing the input poset in a bottom-down fashion. As mentioned above the number of nodes in $\hierarchy$ can be prohibitively large, therefore we also {\em remove} any {\em bad actions} from the running set of actions (Ln.  10-14). 
An action $\alpha$ is bad when $r(\alpha) + \sigma(\alpha) < \max_{\alpha^{\prime} \in \mathcal{S}} (r(\alpha^{\prime}) - \sigma(\alpha^{\prime}))$. Intuitively, this states that we do not need to consider an action as long as there exists another action such that the upper-bounded return of the former is lower than the lower bounded return of the latter. This is a standard technique adopted in multi-armed bandits to limit the number of actions considered by the algorithm~\cite{EvenDar06actionelimination}.

\begin{algorithm}[h]
\small\caption{UpdateActionSet}
\label{algo:updateactions}
\begin{algorithmic}[1]
\STATE {\bf Input:} $\hierarchy$: the hierarchy describing the entity domain; $u$: a node in $\hierarchy$ associated with the last selected action; $\mathcal{S}_{old}$: the running set of actions; $V_k$: set of values for query parameter $k$; $V_l$: set of values for query parameter $l$;
\STATE {\bf Output:} $\mathcal{S}_{new}$: the updated set of actions;
\STATE \textbf{/* Extend Set of Actions*/}
\IF {$\mathcal{S}_{old}$ is empty}
	\RETURN $\{$Root of $\hierarchy \}$
\ENDIF 
\STATE $\mathcal{S}_{new} \leftarrow \mathcal{S}_{old}$
\FORALL{$d \in ${\sf ~Set of Direct Descendant Nodes of $u$ in $\hierarchy$}}
\STATE $A_d \leftarrow$ Set of queries at $u$ for all configurations in $V_k \times V_l$
\STATE $\mathcal{S}_{new} \leftarrow \mathcal{S}_{new} \cup A_d$
\ENDFOR
\STATE \textbf{/* Remove Bad Actions*/}
\STATE /* Find maximum lower bound on gain over all actions in $\mathcal{S}_{new}$*/
\STATE $thres \leftarrow \max_{\alpha^{\prime} \in \mathcal{S}_{new}} (r(\alpha^{\prime}) - \sigma(\alpha^{\prime}))$  
\STATE $\mathcal{B} \leftarrow$ All actions $a$ in $\mathcal{S}_{new}$ with $r(\alpha) + \sigma(\alpha) < thres$
\STATE $\mathcal{S}_{new} \leftarrow \mathcal{S}_{new} \setminus \mathcal{B}$
\RETURN $\mathcal{S}_{new}$
\end{algorithmic}
\end{algorithm}
\vspace{-5pt}

%% file: sections/exps.tex

\section{Experimental Evaluation}
\label{sec:exps}
We present an empirical evaluation of our proposed algorithmic framework using both real and synthetic datasets. First, we discuss the experimental methodology, then we describe the data and results that demonstrate the effectiveness of our framework on crowdsourced entity extraction. The evaluation is performed on an Intel(R) Core(TM) i7 3.7 GHz 32GB machine; all algorithms are implemented in Python 2.7. 

\subsection{Experimental Setup}
\label{sec:expsetup}
\vspace{2pt}\noindent\textbf{Gain Estimators.} We evaluate the following gain estimators:
\squishlist
\item Chao92Shen: This estimator combines the methodology proposed by Chao~\cite{chao:1992} for estimating the number of unseen species  with Shen's formula, i.e., Equation~(1).
\item HwangShen: This estimator combines the regression-based approach proposed by Hwang and Shen~\cite{hwang:2010} for estimating the number of unseen species with Shen's formula. 
\item NewRegr: This estimator corresponds to our new technique proposed in Section~\ref{sec:newestim}.
\squishend
All estimators were coupled with bootstrapping to estimate their variance to retrieve an upper bound on the return of a query as shown in Section~\ref{sec:balancing}.

\vspace{2pt}\noindent\textbf{Entity Extraction Algorithms.} We evaluate the following algorithms for crowdsourced entity extraction:
\squishlist
\item Rand: This algorithm executes random queries until all the available budget is used. It selects a random node from the input poset $\hierarchy$ and a random query configuration $(k,l)$ from a list of pre-specified $k$, $l$ value combinations. \iftr We expect Rand to be effective for extracting entities in small and dense data domains that do not have many sparsely populated nodes. \fi
\item RandL: Same as Rand but only executes queries {\em only at the lowest level nodes} (i.e., leaf nodes) of the input poset $\hierarchy$ until all the available budget is used.  \iftr We expect RandL to be effective for {\em shallow} data domains when the majority of nodes corresponds to leaf nodes. Like Rand, the performance of RandL is expected to be reasonable for small and dense data domains without sparsely populated nodes.\fi
\item BFS: This algorithm performs a breadth-first traversal of the input poset $\hierarchy$, executing one query at each node. The query configuration is randomly selected from a list of pre-specified $k$, $l$ value combinations. This algorithm promotes exploration of the action space when extracting entities. \iftr It also takes into account the structure of the input domain but is agnostic to sparsely populated nodes of the input $\hierarchy$. \fi
\item RootChao: This algorithm corresponds to the entity extraction scheme of Trushkowsky et al.~\cite{trushkowsky:2013} that utilizes the Chao92Shen estimator to measure the gain of an additional query. The proposed scheme is agnostic to the structure of the input entity domain, and thus, equivalent to issuing queries only at the root node of the poset $\hierarchy$. Since the authors only propose a pay-as-you-go scheme, we coupled this algorithm with Alg.~\ref{algo:overall} to optimize for the input budget constraint. We allowed the algorithm to consider different query configurations $(k,l)$ but restricted the possible queries to the root node.
\item GSChao, GSHWang, GSNewR: These algorithms correspond to our proposed querying policy algorithm (Section~\ref{sec:heuristic}) coupled with Chao92Shen, HwangShen and NewRegr respectively.
\item GSExact: This algorithm is used as a near-optimal, omniscient baseline that allows us to see how far off our algorithms are from an algorithm with perfect information. In particular, we combine the algorithm proposed in Section~\ref{sec:heuristic} with an exact computation of the return or gains from queries. More precisely, the algorithm proceeds as follows: At each round we speculatively execute each of the available actions (i.e., all query configurations across all nodes) and select the one that results in the largest number of return to cost ratio. Since the return of each query is known, the algorithm is not coupled with any of the aforementioned estimators.
\squishend

Rand, RandL and BFS promote the exploration of the action space when extracting entities, while the other algorithms balance exploration with exploitation. For the results reported below, we run each algorithm ten times and report the average gain achieved under the given budget.

\vspace{2pt}\noindent\textbf{Querying Interface.} For all datasets we consider generalized queries of the type ``Give me $k$ more entities that satisfy certain conditions and are not present in an exclude list of size $l$''. The conditions correspond to matching the attribute values associated with a node from the input poset. The  configurations considered for $(k,l)$ are {\small $\{(5,0), (10,0), (20,0), (5,2), (10,5), (20,5), (20,10)\}$}. Larger values of $k$ or $l$ were deemed unreasonable for crowdsourced queries. The gain of a query is computed as the number of new entities extracted. The cost of each query is computed using an additive model comprised by three partial cost terms that depend on the characteristics of the query. 

The three partial cost terms are: (i) {\sf CostK} that depends on the number of responses $k$ requested from a user, (ii) {\sf CostL} that depends on the size of the exclude list $l$ used in the query, and (iii) {\sf CostSpec} that depends on the {\em specificity} of the query $q_s$, e.g., we assume that queries that require users to provide more specialized entities (e.g., ``Give me one concert for New York on the 17th of Nov'') cost more than more generic queries (e.g.,  ``Give me one concert in New York''). More formally, we define the specificity of a query to be equal to the number of attributes assigned non-wildcard values for the node $u \in \hierarchy$ the query corresponds to. 

The overall cost for a query with configuration $(k,l)$ with specificity $s$ is computed as:
{\scriptsize 
\begin{equation}
Cost(q) = \alpha \cdot \frac{k}{\mbox{max. query size}} + \beta \cdot  \frac{l}{\mbox{max. ex. list size}} + \gamma \cdot  \frac{s}{\mbox{max. specificity}}
\end{equation}}

The cost of a query should be significantly increased when an exclude list is used, thus we require that $\beta$ is set to a larger value than $\alpha$ and $\gamma$. For the results reported below, we set $\alpha = \gamma = 1$ and $\beta = 5$. Similar results were observed for other settings.

\vspace{2pt}\noindent\textbf{Data.} First, we evaluate the proposed framework on extracting entities from a large sparse domain. We consider the event dataset collected from Eventbrite. As described in Section~\ref{sec:intro}, the poset corresponding to the Eventbrite domain contains 8,508,160 nodes with 57,805 distinct events overall. However, only 175,068 nodes are populated leading to a rather sparsely populated domain. Due to lack of popularity proxies for the extracted events, we assigned a random {\em popularity value} in $(0,10]$ to each event. These weights are used during sampling to form the actual popularity distribution characterizing the population of each node in the poset. 

We further evaluate the performance of the extraction algorithms for a more dense domain, that we constructed ourselves. We used Amazon's Mechanical Turk~\cite{mturk} to collect a real-world dataset, targeted at extracting ``people in the news''. While different from the event extraction domain studied before this new domain is still structured. We asked workers to extract the names of people belonging to four different types from five different news portals. The people types we considered are ``Politicians'', ``Athletes'', ``Actors/Singers'' and ``Industry People''. The news portals we considered are ``New York Times'', ``Huffington Post'', ``Washington Post'', ``USA  Today'' and ``The Wall Street Journal''. This data domain, referred to as the People's domain, is essentially characterized by the type of the individual and the news portal. Workers were paid \$0.20 per HIT. We issued 20 HITS for each leaf node of the domain's poset, resulting in 600 HITS in total. After manually curating name misspelling's, we extracted 1,245 unique people in total. Table~\ref{tab:ptypedata} shows the number of distinct entities for the different values of the people-type and news portal attributes. Finally, the popularity value of each extracted entity was assigned to be equal to the number of times it appeared in the extraction result. The values are normalized during sampling time to form a proper popularity distribution. Collecting a large amount of data in advance from Mechanical Turk and then simulating the responses of human workers by revealing portions of this dataset allows us to compare different algorithms on an equal footing; this approach is often adopted in the evaluation of crowdsourcing algorithms~\cite{DBLP:journals/pvldb/ParameswaranBG0PW14, marcus:2011,trushkowsky:2013}.

\begin{table}
\scriptsize\center
\caption{The population characteristics for the People's domain.}
\label{tab:ptypedata}
\begin{tabular}{|c|c|}
\hline
\textbf{Person Type} & \textbf{People} \\ \hline
Industry People & 743 \\
Athletes & 743 \\
Politicians & 748 \\
Actors/Singers & 744 \\ \hline
\end{tabular}
\quad
\begin{tabular}{|c|c|}
\hline
\textbf{News Portal} & \textbf{People} \\ \hline
WSJ & 594 \\
WashPost & 597 \\
NY Times & 595 \\
HuffPost & 599 \\
USA Today & 593 \\ \hline
\end{tabular}
\vspace{-15pt}
\end{table}

\subsection{Experimental Results}
Next, we evaluate different aspects of the aforementioned extraction techniques. 

\begin{figure}[h]
\begin{center}
        \subfigure{\includegraphics[trim=120 0 0 0,scale=0.40]{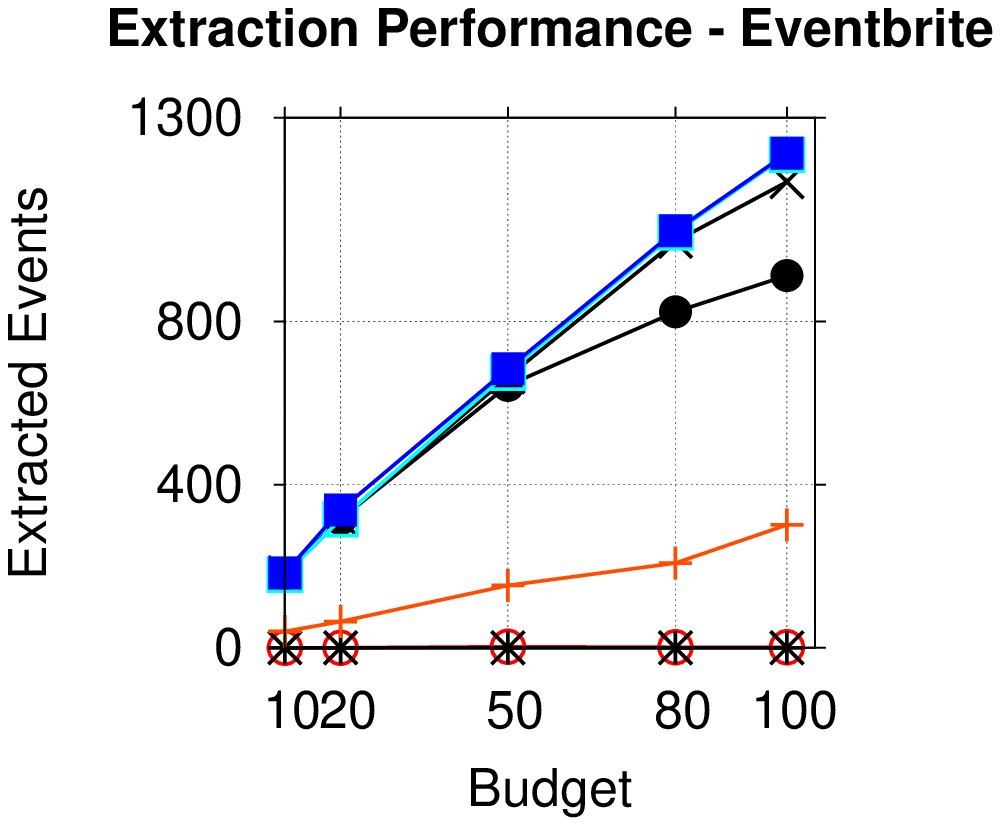} \label{fig:ebextraction}}
        \subfigure{\includegraphics[trim=80 0 100 0,scale=0.40]{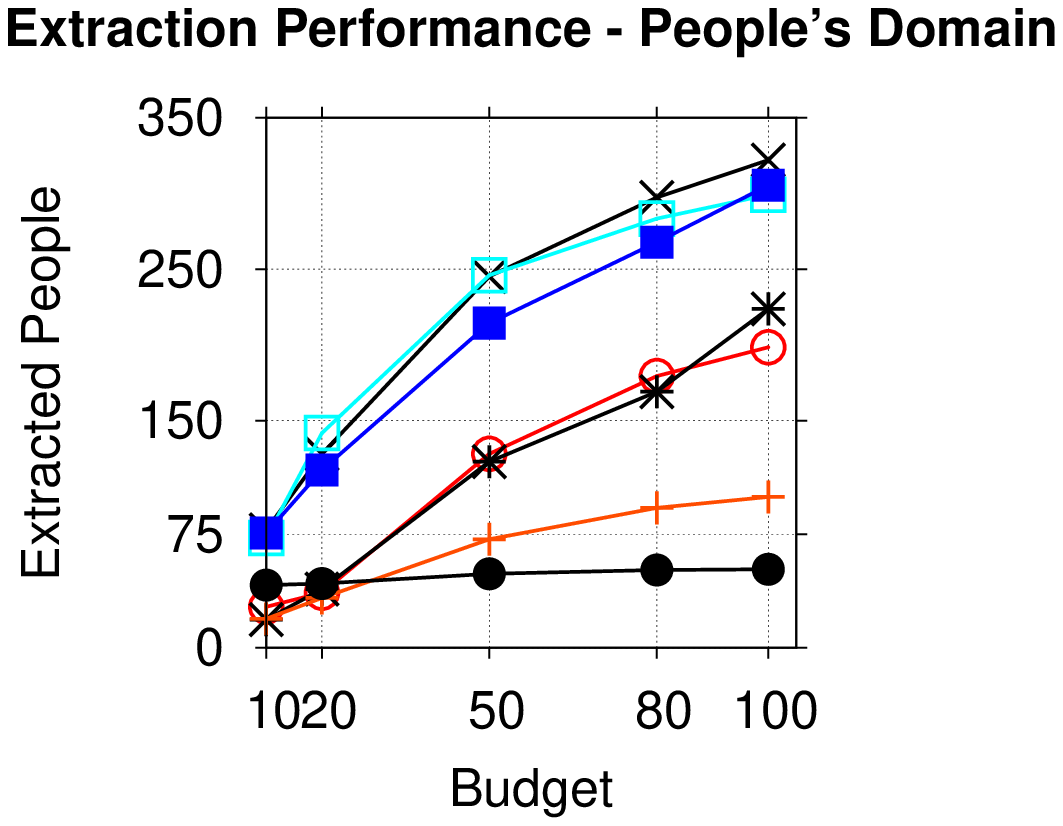} \label{fig:poextraction}}
        \subfigure{\includegraphics[scale=0.30]{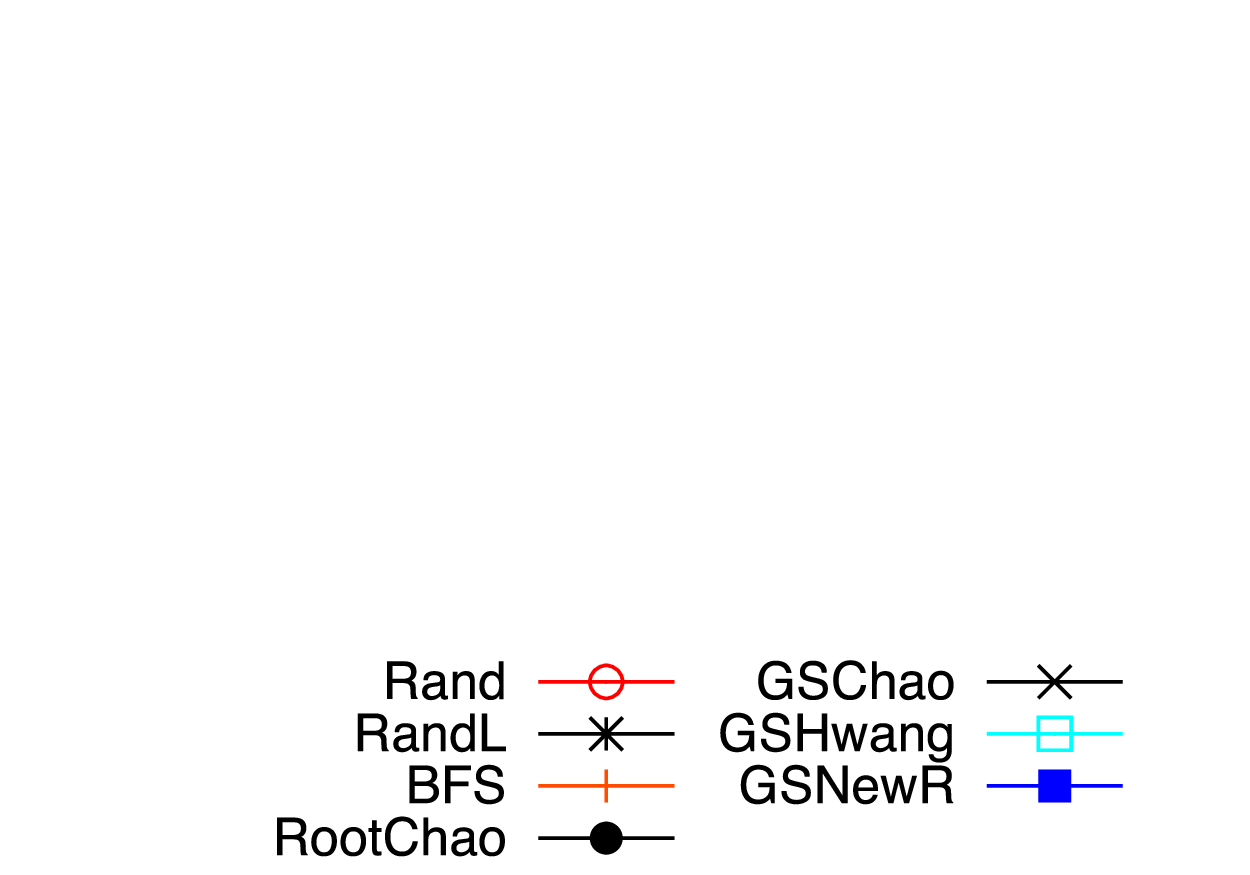}}
\end{center}
\vspace{-20pt}
\caption{A comparison of the proposed entity extraction techniques against several baselines for (a) Eventbrite and (b) the People's domain.}
\label{fig:resultsextr}
\vspace{-15pt}
\end{figure}

\vspace{3pt}\noindent{\ul{\textbf{How does our querying policy algorithm compare against baselines?}}}
We evaluate the performance of the different extraction algorithms in terms of number of entities extracted for different budgets. The results for Eventbrite and the People's domain are shown in Figure~\ref{fig:ebextraction} and Figure~\ref{fig:poextraction} respectively. As shown, our proposed algorithms, i.e., GSChao, GSHwang, GSNewR {\em outperform all baselines for at least 30\% across both datasets}. This behavior is expected as our techniques not only exploit the structure of the domain to diversify entity extraction by targeting entities that belong to the tail of the popularity distribution but also optimize the queries for the given budget.

When comparing again the naive baselines Rand, RandL, and BFS, we see that GSChao, GSHwang and GSNewR extract at least 2X more entities for the sparse Eventbrite domain and around 100\% more entities for small budgets and 54\% for larger ones when considering the dense People's domain. For example for Eventibrite and a budget of \$50 all schemes coupled with our querying policy discovery algorithm (Section~\ref{sec:solving}) extracted more than 600 events while Rand and RandL extracted 1.1 and 0.2 events and BFS extracted 207.7 events, an improvement of over 180\%.

Comparing against RootChao, we see that GSChao, GSHwang and GSNewR, are able to retrieve up to 30\% more entities for Eventbrite and 5X for the People's domain. This performance difference is due to the fact that the gain achieved by RootChao saturates at a faster rate compared to GSChao, GSHwang and GSNewR as the cost increases. This is because, RootChao focuses on issuing queries at the root of the input poset, and hence, it is not able to extract entities belonging to the long tail of the popularity distribution. Moreover, for the People's domain we see that RootChao performs poorly even compared to the naive baselines Rand, RandL and BFS. Again, this behavior is due to the skew of the underlying popularity distribution.

\begin{figure}[h]
\vspace{-10pt}
\begin{center}
        \subfigure{\includegraphics[trim=120 0 0 0,scale=0.40]{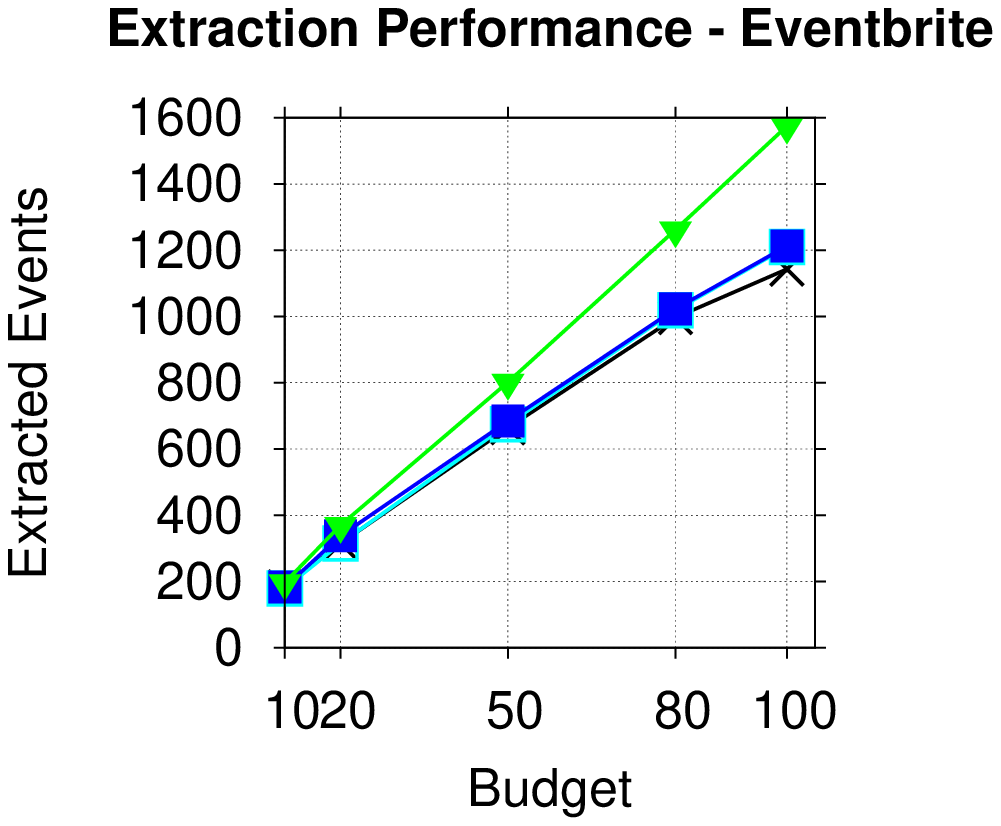} \label{fig:ebextractionopt}}
        \subfigure{\includegraphics[trim=80 0 100 0,scale=0.40]{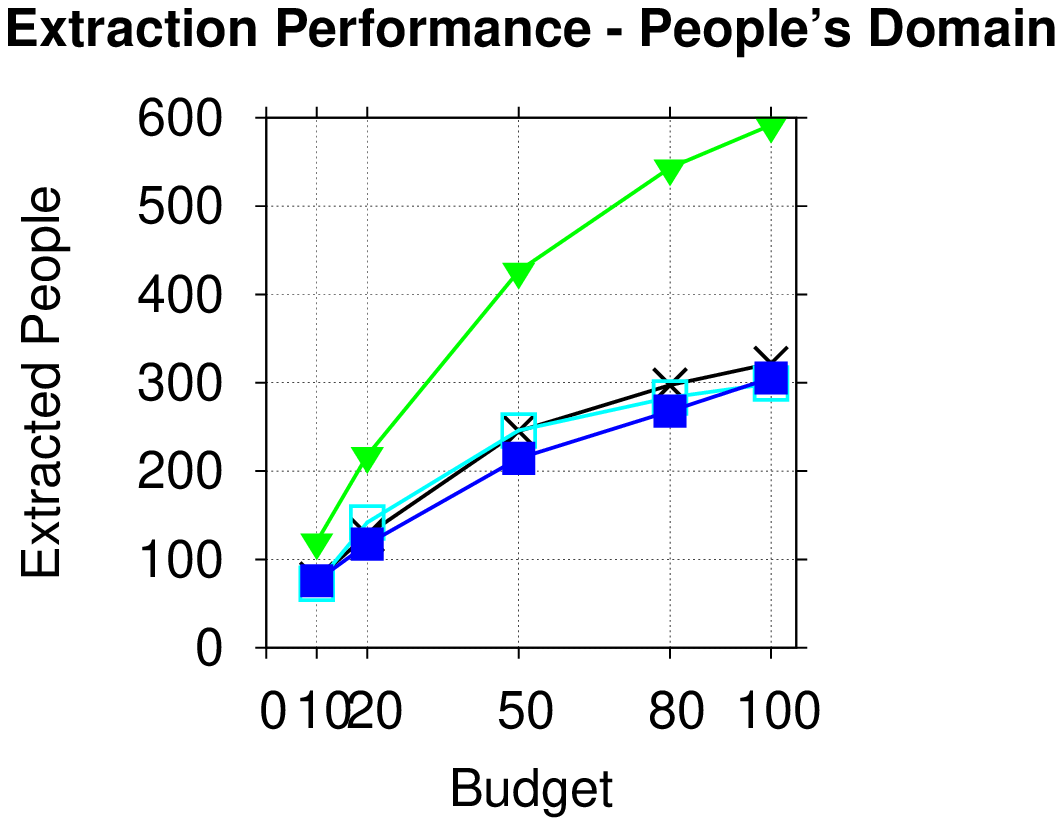} \label{fig:poextractionopt}}
        \subfigure{\includegraphics[scale=0.30]{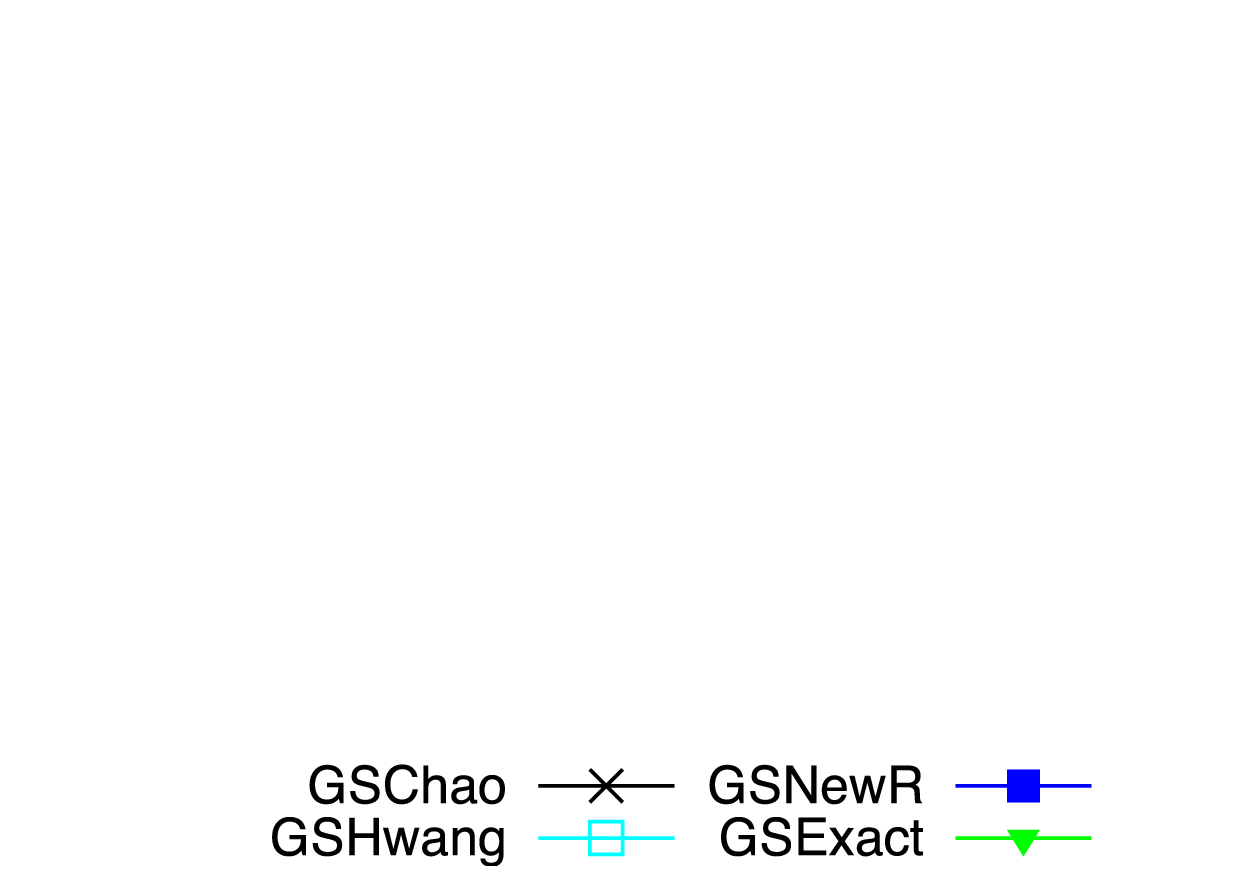}}
\end{center}
\vspace{-20pt}
\caption{A comparison of the proposed entity extraction techniques against a near-optimal algorithm for (a) Eventbrite and (b) the People's domain.}
\label{fig:resultsextr}
\vspace{-10pt}
\end{figure}

\vspace{3pt}\noindent{\ul{\textbf{How do our techniques compare against a near-optimal policy discovery algorithm?}}}
Next, we evaluate GSChao, GSHwang and GSNewR against the near-optimal querying policy discovery algorithm GSExact. The results for Eventbrite and the People's domain are shown in Figure~\ref{fig:ebextractionopt} and Figure~\ref{fig:poextractionopt} respectively. Regarding the dense domain Eventbrite, we observe that for smaller budgets our proposed techniques perform comparably to GSExact that has ``perfect information'' about the gain of each query, typically demonstrating a performance gap of less than 10\%. For larger budgets this gap increases to 25\%. Note that our estimators have access to few samples and sparse information; the fact that we are able to get this close to GSExact is notable. Finally, for the People's domain, our techniques present an increased performance gap compared to GSExact. Nevertheless the performance drop is at most 50\%. 

\begin{figure}[h]
	\begin{center}
	\includegraphics[clip,scale=0.4]{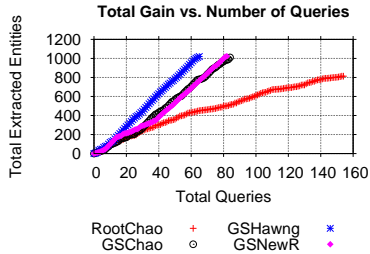}
	\caption{The number of events extracted by different algorithms for the Eventbrite data domain and the corresponding total number of queries.}
	\label{fig:rounds}
	\end{center}
\end{figure}

\vspace{3pt}\noindent{\ul{\textbf{How do the different techniques compare with respect to the total number of queries issued during extraction?}}}
We compare the performance of RootChao (i.e., the extraction scheme proposed by Trushkowsky et al.~\cite{trushkowsky:2013}) against our algorithms GSChao, GSHwang and GSNewR with respect to the {\em total number of queries issued during extraction}. Notice that this new evaluation metric characterizes directly the overall latency of the crowd-extraction process. Figure~\ref{fig:rounds} shows the corresponding results for a run for Eventbrite and a budget of \$80. As shown {\em RootChao requires almost up to 3x more queries} to extract the same number of entities as our proposed techniques, thus, exhibiting significantly larger latency compared to GSChao, GSHwang and GSNewR.

\begin{figure}[h]
	\vspace{-10pt}
    \includegraphics[clip,scale=0.32]{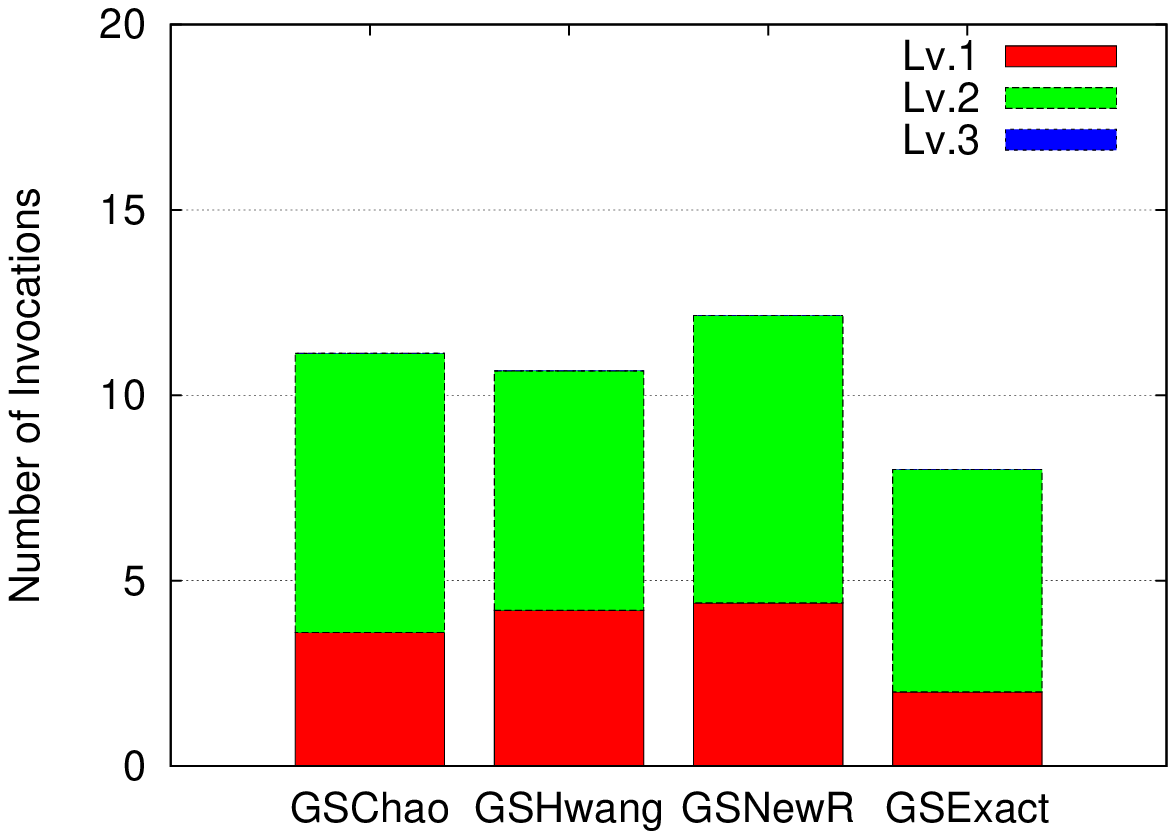}
	\hspace{-10pt}
	\includegraphics[clip,scale=0.32]{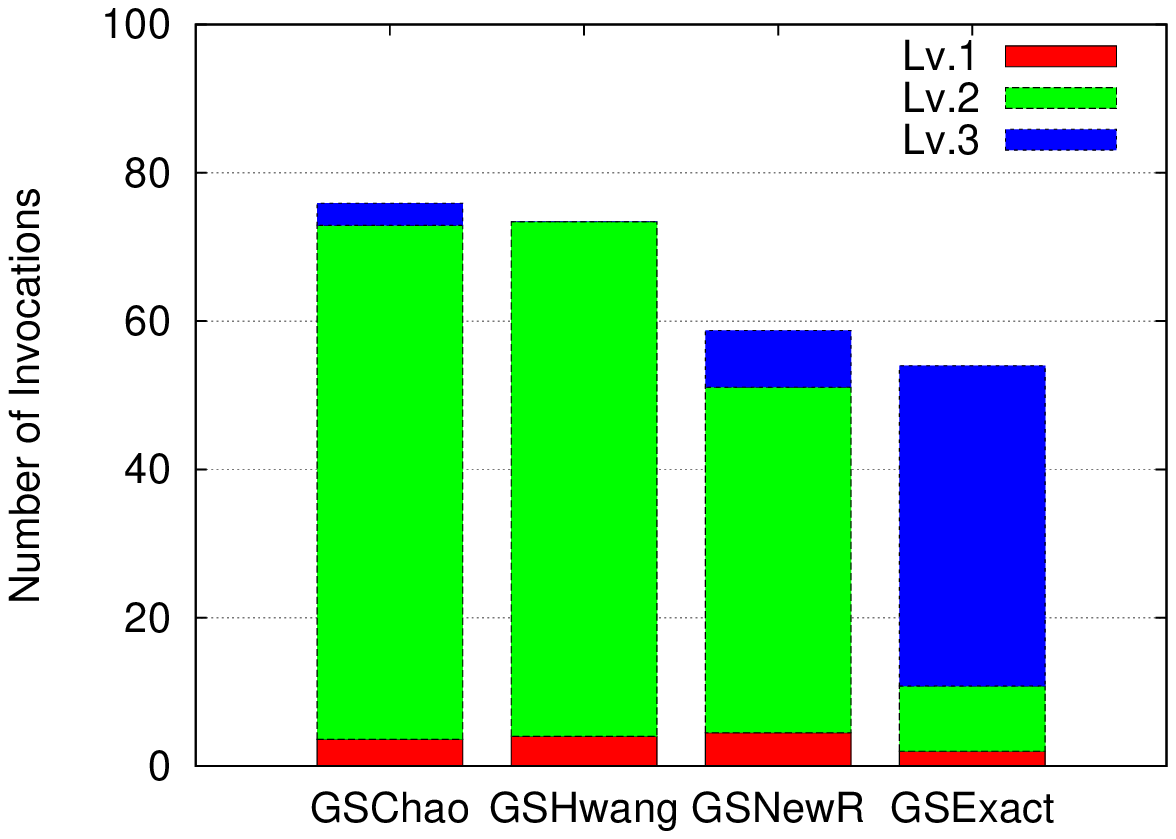}
	\vspace{-10pt}
	\caption{The number of queries issued at different levels used when budget is set at 10 or 100.}\label{fig:level}
	\vspace{-10pt}
\end{figure}

\begin{figure}[h]
	\vspace{-5pt}
   	 \includegraphics[clip,scale=0.32]{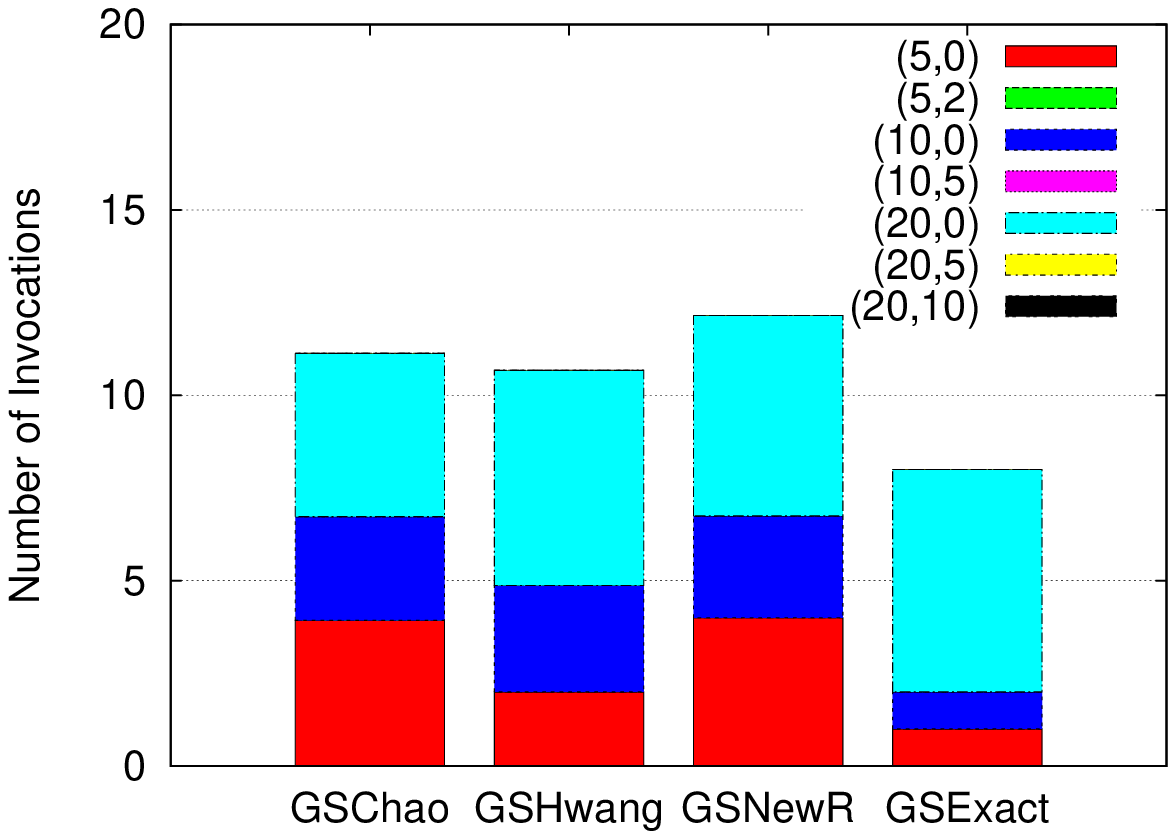}
	\hspace{-10pt}
	\includegraphics[clip,scale=0.32]{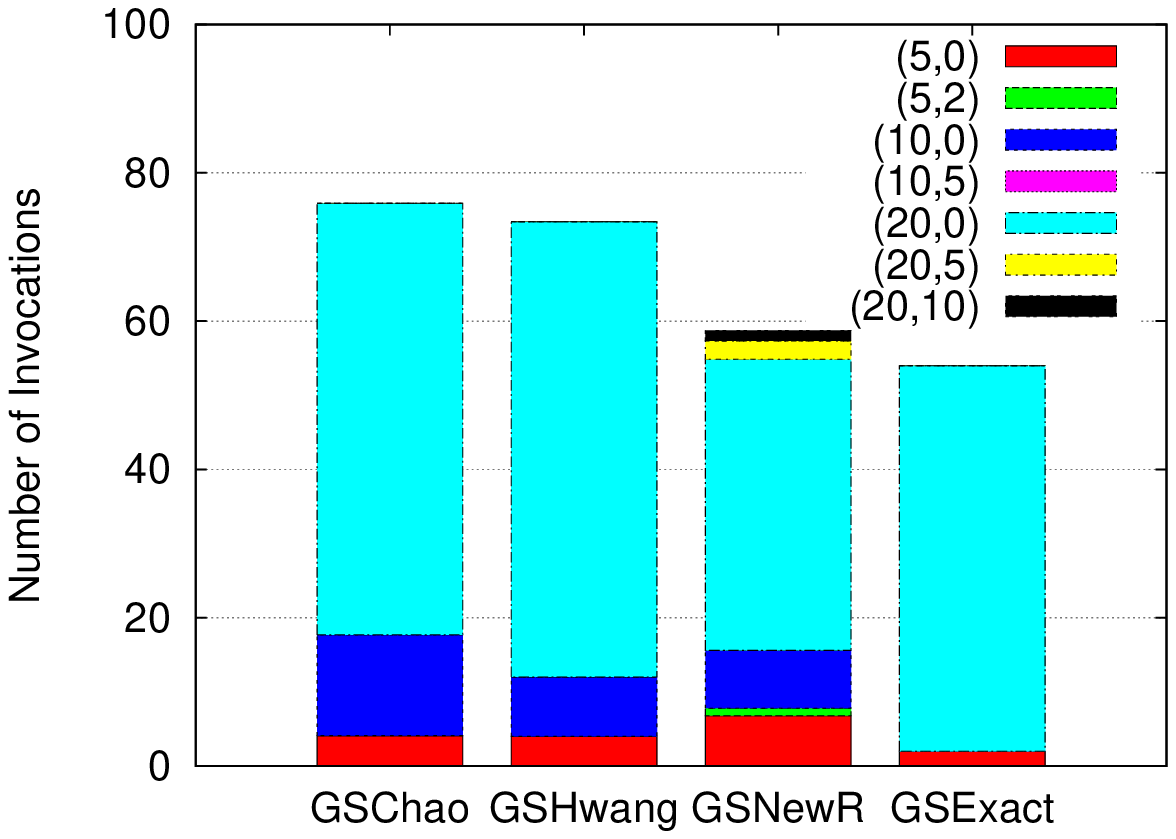}
	\vspace{-10pt}
	\caption{The query configurations used when budget is set at 10 or 100.}\label{fig:queryconf}
	\vspace{-10pt}
\end{figure}

\vspace{3pt}\noindent{\ul{\textbf{How our different algorithms traverse the poset and use different query configurations?}}}
We next explore how our different algorithms traverse the poset, and how they use different query configurations. The results reported are averaged over ten runs and correspond to the People's domain. We begin by considering how many queries these algorithms issue at various levels of the poset. In Figure~\ref{fig:level}, we plot the different number of queries issued at various levels by our algorithms when the budget is set to 10 and 100 respectively. Given a small budget, we observe that all algorithms prefer issuing queries at higher levels of the poset. Notice that inner nodes of the poset are preferred and only a small number of queries is issued at the root (i.e., level one) of the poset. This behavior is justified if we consider that due to their popularity, certain entities are repeatedly extracted, thus leading to a lower gain. As the budget increases, we see that all algorithms tend to consider more specialized queries at deeper levels of the poset. It is interesting to observe that all of our algorithms issue the majority of their queries at the level two nodes, while GSExact, which has perfect information, focuses mostly on the leaf nodes. Thus, in this case, our techniques could benefit from being more aggressive at traversing the poset and reaching deeper levels; overall, our techniques may end up being more conservative in order to cater to a larger space of posets and popularity distributions. In Figure~\ref{fig:queryconf}, we plot the different query configurations chosen by our algorithms when the budget is set to 10 and 100 respectively. We observe that GSExact always prefers queries with $k = 20$ and $l = 0$ for both small and large budgets. On the other hand, our algorithms issue more queries of smaller size when operating under a limited budget and prefer queries of larger size for larger budgets. Out of all algorithms we see that GSNewR was the only one issuing queries with exclude lists of different sizes, thus exploiting the rich diversity of query interfaces. However, the number of such queries is limited.

\vspace{3pt}\noindent{\ul{\textbf{How effective are the different estimators at predicting the gain of additional queries?}}}
Finally, we point out that GSNewR was able to outperform GSChao and GSHwang for Eventbrite but the opposite behavior was observed for the People's domain. To further understand the relative performance of GSChao, GSHwang and GSNewR, we evaluate the performance of the gain estimators Chao92Shen, HwangShen and NewRegr at predicting the number of new retrieved events for different query configurations. For Eventbrite, we choose ten random nodes containing more than 5,000 events and for each of them and each of the available query parameter configurations $(k,l)$, we execute ten queries of the form ``Give me $k$ items from node $u \in \hierarchy$ that are not included in an exclude list of size $l$''. As mentioned in Section~\ref{sec:excludelist} the exclude list for each query is constructed following a randomized approach.  For the People's domain, we issue ten queries over all nodes of the input poset for all available query configurations.  We measure the performance of each estimator by considering the absolute relative error between the predicted return and the actual return of the query. 

Table~\ref{tab:eventesterror} reports the relative error for each of the three estimators averaged over all points under consideration for Eventbrite. As shown, all three estimators perform equivalently with the new regression-based technique slightly outperforming Chao92Shen and HwangShen for certain types of queries. For example, for $k = 10, l = 5$, Chao92Shen has a relative error of 0.58, HwangShen had a relative error of 0.7, and NewRegr had a relative error of 0.29. We attribute the improved extraction performance of GSNewR to these improved estimates. The relatively large values for relative errors are justified as the retrieved samples correspond to a very small portion of the underlying population for each of the points. This is a well-known behavior for non-parametric estimators and studied extensively in the species estimation literature~\cite{hwang:2010}. 

\begin{table}
\scriptsize\center
\caption{Average absolute relative error for estimating the gain of different queries for Eventbrite.}
\label{tab:eventesterror}
\begin{tabular}{|c|c|c|c|c|}
\hline
\textbf{Q. Size $k$} & \textbf{EL. Size $l$} & \textbf{Chao92Shen} & \textbf{HwangShen} & \textbf{NewRegr} \\ \hline
5 & 0 & 0.470 & 0.500 & 0.390 \\
5 & 2 & 0.554 & 0.612 & 0.467\\
10 & 0 & 0.569 & 0.592 & 0.544\\
10 & 5 & 0.580 & 0.696 & 0.29\\
20 & 0 & 0.642 & 0.756 &0.471\\
20 & 5 & 0.510 & 0.60 & 0.436 \\
20 & 10 & 0.653 & 0.756 & 0.631\\
\hline
\end{tabular}
\vspace{-10pt}
\end{table}

Table~\ref{tab:peopleesterror} shows the results for the People's domain. We observe that for smaller query sizes the regression technique proposed in this paper offers better gain estimates. However, as the query size increases, and hence, a larger portion of the underlying population is observed Chao92Shen outperforms both regression-based techniques. Thus, we are able to explain the performance difference between GSChao and the other two algorithms. Eventually, we have that for sparse domains regression-based techniques result in better performance. However, for dense domains the Chao92Shen estimator results in better performance as a larger portion of the underlying population can be sampled. 

\begin{table}[h]
\vspace{-5pt}
\scriptsize \center
\caption{Average absolute percentage error for estimating the gain of different queries for the People's data domain.}
\vspace{-5pt}
\label{tab:peopleesterror}
\begin{tabular}{|c|c|c|c|c|}
\hline
\textbf{Q. Size $k$} & \textbf{EL. Size $l$} & \textbf{Chao92Shen} & \textbf{HwangShen} & \textbf{NewRegr} \\ \hline
5 & 0 & 0.295 & 0.299 & 0.228\\
5 & 2 & 0.163 &  0.156 & 0.144\\
10 & 0 &  0.306 & 0.305 & 0.277\\
10 & 5 &  0.341 & 0.349 & 0.293\\
20 & 0 &  0.359& 0.371 & 0.467 \\
20 & 5 &  0.2615 & 0.264 & 0.249\\
20 & 10 & 0.1721 & 0.162 & 0.127\\
\hline
\end{tabular}
\vspace{-15pt}
\end{table}

%% file: sections/related.tex

\section{Related Work}
\label{sec:related}
The prior work related to the techniques proposed in this paper can be placed in a few categories; we describe each of them in turn:

\vspace{3pt}\noindent\textbf{Crowd Algorithms.} There has been a significant amount of work on designing algorithms where the unit operations (e.g., comparisons, predicate evaluations, and so on) are performed by human workers, including common database primitives such as filter~\cite{crowdscreen}, join~\cite{markus-sorts-joins} and max~\cite{so-who-won},  machine learning primitives such as entity resolution~\cite{entity-matching, crowder} and clustering~\cite{crowdclustering}, as well as data mining primitives~\cite{amsterdamer:2013, get-another-label}. 

Previous work on the task of crowdsourced extraction or enumeration, i.e., populating a database with entities using the crowd~\cite{park:2014, trushkowsky:2013} is the most related to ours. In both cases, the focus is on a single entity extraction query; extracting entities from large and diverse data domains is not considered. Moreover, the proposed techniques do not support  dynamic adaptation of the queries issued against the crowd to optimize for a specified monetary budget. 


\vspace{3pt}\noindent\textbf{Knowledge Acquisition Systems.} Recent work has also considered the problem of using crowdsourcing within knowledge acquisition systems~\cite{jiang:13, kondredi:2014, west:2014}. This line of work suggests using the crowd for curating knowledge bases (e.g., assessing the validity of the extracted facts) and for gathering additional information to be added to the knowledge base (e.g., missing attributes of an entity or relationships between entities), instead of augmenting the set of entities themselves. As a result, these papers are solving an orthogonal problem. The techniques described in this paper for estimating the amount of information from a query and devising querying strategies to maximize the amount of extracted information will surely be beneficial for knowledge extraction systems as well.

\vspace{3pt}\noindent\textbf{Deep Web Crawling.} A different line of work has focused on data extraction from the deep web~\cite{Jin:2011,Sheng:2012}. In such scenarios, data is obtained by querying a form-based interface over a hidden database and extracting results from the resulting dynamically-generated answer (often a list of entities). Typically, such interfaces provide partial list of matching entities to issued queries; the list is usually limited to the top-k tuples based on an unknown ranking function. Sheng et al.~\cite{Sheng:2012} provide near-optimal algorithms that exploit the exposed structure of the underlying domain to extract all the tuples present in the hidden database under consideration. Our work is similar to this work in that our goal is to also extract entities via a collection of interfaces (in our case the interfaces correspond to queries asked to the crowd).

The main difference between this line of work and ours is that answers from a hidden database are deterministic, i.e., a query in their setting will always retrieve the same top-k tuples. This assumption does not hold in the crowdsourcing scenario considered in this paper and thus the proposed techniques are not applicable. In their setting, it suffices to ask each query precisely once. In our setting, since crowdsourced entity extraction queries can be viewed as random samples from an unknown  distribution, one needs to make use of the query result estimation techniques introduced in Section~\ref{sec:gainestimators}.

%% file: sections/conclusions.tex

\section{Conclusions and Future Work}
\label{sec:conclusions}
In this paper, we studied the problem of crowdsourced entity extraction over large and diverse data domains. We introduced a novel crowdsourced entity extraction framework that combines statistical techniques with an adaptive optimization algorithm to maximize the total number of unique entities extracted. We proposed a new regression-based technique for estimating the gain of further querying when the number of retrieved entities is small with respect to the total size of the underlying population. We also introduced a new algorithm that exploits the often known structure of the underlying data domain to devise adaptive querying strategies. Our experimental results show that our techniques extract up to 4X more entities compared to a collection of baselines, and for large sparse entity domains are at most 25\% away from an omniscient adaptive querying strategy with perfect information.

Some of the future directions for extending this work include reasoning about the quality and correctness of the extracted result as well as extending the proposed techniques to other types of information extraction tasks. As mentioned before, the techniques proposed in this paper do not deal with incomplete and imprecise information. However, there has been an increasing amount of literature on addressing these quality issues in crowdsourcing~\cite{ vox-populii, quality, nushi:14, raykar-whom-to-trust}. Combining these techniques, or entity resolution techniques~\cite{crowder} that reason about similarity of extracted entities, with our proposed framework is a promising future direction. Finally, it is of particular interest to consider how the proposed framework can be applied to other budget sensitive information extraction applications including discovering valuable data sources for integration tasks~\cite{rekatsinas:2015, rekatsinas:2014} or curating and completing a knowledge base~\cite{kondredi:2014}.